\documentclass[12pt,preprint]{aastex}

\shorttitle{Radio studies of SN 1993J with GMRT}
\shortauthors{P. Chandra et al.}

\begin{document}

\title{The late time radio emission from SN 1993J at meter wavelengths}

\author{P. Chandra}
\affil{Tata Institute of Fundamental Research,
Mumbai 400 005, India; {\tt poonam@tifr.res.in}}
\affil{Joint Astronomy Programme, Indian Institute of Science,
    Bangalore 560 012, India}

\author{A. Ray}
\affil{Tata Institute of Fundamental Research, Mumbai
400 005, India; {\tt akr@tifr.res.in}}
\and
\author{S. Bhatnagar}
\affil{National Radio Astronomy Observatory, Socorro, NM 87801;
{\tt sbhatnag@aoc.nrao.edu}}

\begin{abstract}

This paper presents the investigations of SN 1993J using  low
frequency observations with the Giant Meterwave Radio Telescope. We
analyze the light curves of SN 1993J at 1420, 610, 325 and 243 MHz
during  $7.5-10$ years since explosion. 
The supernova has become optically thin early on in the 1420 MHz and 610 MHz
bands while it has only recently entered the optically thin phase in
the 325 MHz band. The radio light curve in the 235 MHz band is more 
or less flat. This indicates that the supernova is undergoing a transition
from an optically thick to optically thin limit in this frequency band.
In addition,  we analyze the SN radio 
spectra at five epochs on day 3000, 3200, 3266, 3460
and 3730 since explosion. 
SN 1993J is the only young supernova for which the magnetic
field and the size of the radio emitting region are determined
through unrelated methods independently.
Thus the mechanism that controls the evolution of the
radio spectra can be identified.
We suggest that at all epochs, the synchrotron self
absorption mechanism is primarily responsible for the turn-over in the
spectra.  
Light curve models based on  
free free absorption in homogeneous or inhomogeneous media  at
high frequencies overpredict
the flux densities at low frequencies. The discrepancy is increasingly
larger at 
lower and lower frequencies. We suggest that an extra opacity, 
sensitively dependent on frequency, is likely to
account for the difference at lower frequencies.  
The evolution of the magnetic field (determined from synchrotron
self absorption turn-over)
is roughly consistent with $B \propto t^{-1}$.
Radio spectral index in the optically thin part 
evolves from $\alpha \sim 0.8-1.0$ at few tens of days to 
$\sim 0.6$ in about ten years.


\end{abstract}


\section{Introduction}

Supernova SN 1993J exploded on March 28, 1993 \citep{rip93} 
in the nearby galaxy M81
at a distance of 3.6 Mpc \citep{fre94}. 
The early spectrum of SN 1993J showed the
characteristic hydrogen line signature of type II supernovae, 
but subsequently made a
transition to hydrogen-free, helium-dominated type Ib supernova
\citep{swa93,fil93}. SN 1993J was therefore classified as an
archetypal 'type IIb' supernova.
The supernova provided a very good opportunity for a detailed study of
extragalactic supernovae because, first, it was one of the nearest
extragalactic supernova  and secondly, it was easily observable from the
northern hemisphere for 
the most part of a year due to its high positive 
declination. It was extensively observed 
soon after its discovery in the non-optical bands as well. 
The radio detection at 1.4 cm with the VLA took place on day 5
\citep{wei93} while in the X-ray bands, ROSAT detected it on 
day 6 \citep{zim93}.

Optical observations of the pre- and post-supernova fields of 
SN 1993J indicated that its progenitor was an
early K-type supergiant star with $M_V=-7.0 \pm 0.4$ magnitude
and an initial mass 
of $13-22 M_{\odot}$ \citep{van02}. Photometric evolution
of SN 1993J indicated that it had very little hydrogen left 
in the outermost shell. 
This suggested that the progenitor was 
more likely a part of a binary system. The outermost
hydrogen envelope of the progenitor was largely
stripped off by the massive binary companion 
\citep{ray93,nom93,pod93,woo94,utr94}. The possibility that it 
was a  very massive
Wolf-Rayet star ($\sim 30 M_{\odot}$), which lost most of its
outermost envelope before explosion had also been advocated
\citep{hof93}. However, recent high resolution photometric and spectroscopic
observations of SN 1993J with the Hubble Space Telescope, ten years after the
explosion
has  unambiguously detected the signature of the
massive binary companion of SN 1993J with almost the same mass ($14-15 
M_{\odot}$) as that 
of the progenitor of SN 1993J \citep{mau04}.

Radio emission from supernovae is  due to the interaction of the
supernova ejecta with the circumstellar medium (CSM).  CSM is created
in the early evolutionary phases of a progenitor star due to 
the mass loss from the  star before it undergoes explosion.
Supernova ejecta moving with supersonic speeds,
upon interaction with the CSM gives rise to a low density, high
temperature forward shocked shell and a high density, low
temperature reverse shocked region. The radio emission is believed to
be due to non-thermal electron 
synchrotron emission from the hydrodynamically unstable interaction region.
This region contains shocked supernova ejecta as well as
swept-up circumstellar 
material which are separated by a contact discontinuity and is bounded 
outwards by the blast wave shock and inwards by the reverse shock.
The radio emission can be absorbed by the
surrounding dense circumstellar medium through free-free absorption
(FFA). It can also undergo synchrotron self absorption (SSA) in
the magnetized plasma. In general, light curves of
supernovae show an initial smooth rise of 
radio flux density in their early stage
because of decreasing absorption of the overlying medium and 
subsequently  a
smooth fall due to the  decreasing density of the circumstellar medium. 
Optical depth of free-free absorption is defined as $\tau =\int k_{ff} ds 
\propto R^{-3}{\nu}^{-2} $, and $k_{ff}$ is absorption coefficient, which
depends upon  $n_e$ and $n_i$,
electron and ion number densities respectively.
Since $n_e \propto R^{-2}$, $n_i= \overline{Z} n_e$ and thermally averaged
$k_{ff} \propto {\nu}^{-2.1} T^{-1.35}$;  the 
dependence of the optical depth on size and frequency is 
$\tau \propto R^{-3}{\nu}^{-2.1} $.
The light curves
show that the lower frequencies progressively become optically thin at
later epochs, i.e. the radio emission peak shifts to lower frequencies with
time. In some supernovae, FFA is dominant while in
others SSA may become important depending on
the mass loss rate, shock velocity and circumstellar temperature
etc. SSA is usually dominant in ejecta with high velocities and
high CSM temperatures whereas high mass loss rate favors FFA
mechanisms \citep{che03}. For example, the light curves of SN 1979C
fit well with an FFA model \citep{wei91} whereas SN 1998bw
evolution can be well described by an SSA model \citep{kul98}. 
In some cases, the light curves exhibit non-smooth behavior  (e.g., steep
decline or sharp jump in the flux density). This  could be due to
inhomogeneities in the CSM. 
The magnetic field in the shocked shell of a supernova, amplified 
by instabilities near the contact surface, may 
also 
have an inhomogeneous distribution.

Even though SN 1993J has been extensively observed at high radio
frequencies, its evolution at low frequencies is critical
at late epochs. We sampled the light curves of SN 1993J at
lower frequencies between $7.5-10$ years after explosion using the
Giant Meterwave Radio Telescope (GMRT) at 1420, 610, 325 and 235 MHz
bands. We also obtained simultaneous to near-simultaneous spectra of
the SN on five occasions. 
On one occasion, we combined GMRT plus VLA data and obtained a wide-band 
radio spectrum. This composite spectrum
around 3200 days shows an evidence of synchrotron
cooling. The implications of this result 
on the magnetic field and equipartition
fraction in SN 1993J are published elsewhere \citep{cha04}. 
These results are used
extensively in this paper.
Our data additionally suggest that the turn-over in the radio
spectrum can be explained reasonably well by SSA.  FFA appears to have a 
minor role compared to SSA in determining the spectral turn-over at
low frequencies. 
We note that the simple extrapolation of the FFA model obtained 
from high frequency observations of SN 1993J fails to reproduce 
and overpredicts the observed flux densities at low frequencies.
We discuss the evolution of the supernova's 
size, magnetic field and radio spectral index
with time, starting from a few tens of days to 3700 days since
explosion.

Section 2 gives the details of our observations and data analysis. We
discuss in section 3 the light curves and spectra of SN 1993J at low
frequencies. We also give our interpretations of the light curves 
and spectra in Section 3. Summary and conclusions 
are given in Section 4.

\section{Observations and data analysis}

\subsection{GMRT observations}

We observed SN 1993J with the Giant Meterwave Radio Telescope (GMRT)
on several occasions in multiple frequencies (1420, 610, 325 and
235~MHz).  The monitoring program for SN 1993J was started 
in November 2000 and continued till  June 2003. 
We also obtained simultaneous to near-simultaneous spectra of the SN
on five occasions. Total time spent on the supernova during the
observations at various epochs varied from $2-4$ hours. About
17 to 29 good antennas could be used in the radio
interferometric setup for flux density
measurements of the SN at 
different observing epochs. For 1420, 610 and 325 MHz bands
the bandwidth used was 16 MHz (divided in total 128 frequency channels,
-- default for GMRT correlator)
while it was 6 MHz for  243 MHz wave band. Table
\ref{tab:1} gives the observing log for  SN 1993J.

\subsection{Calibrators}

%
%
Calibrator sources were used to remove the effect of variation of 
the instrumental
factors in the measurements. 3C48 and 3C147 were used as flux
calibrators at 1420 MHz.  At lower frequencies i.e. 610, 325 and 243
MHz bands, 3C286, 3C48 and 3C147 were used as flux calibrators.  The flux
densities of the flux calibrators at observing frequencies were 
derived using Baars formulation \citep{bar77}.
 Table \ref{fluxcal} lists the flux calibrators and their flux densities at
 all GMRT wavebands.

For 1420 MHz observations, the source 1035+564 (J2000 co-ordinates:
$10^{\mathrm{h}}35^{\mathrm{m}}07\fs0$,
$+56\arcdeg28\arcmin47\arcsec$) was used as a phase calibrator 
on all occasions except
on 2000 Nov 08, 2000 Dec 16 and 2001 Jun 2 observations when
0834+555 was used as the phase calibrator. For 610, 325 and 235 MHz band
observations, 0834+555 (J2000 co-ordinates
$08^{\mathrm{h}}34^{\mathrm{m}}54\fs9$,
$+55\arcdeg34\arcmin21\arcsec$) was used as a phase calibrator in 
all observations carried out by us.
We used the flux calibrators and 
phase calibrators for bandpass calibration as well. Flux calibrators
were observed once or twice for 20-30 minutes during each observing session.
Phase calibrators were observed for 5-6 minutes after every 25 minutes 
of observation of the supernova. Observations of the phase calibrators
separated by small time intervals were important not only for the tracking of
instrumental phase and gain drifts, atmospheric and ionospheric gain 
and phase variations, but also for monitoring the 
quality and sensitivity of the data and for
spotting occasional gain and phase jumps. Table \ref{1035cal}
gives the details of the observations of phase calibrator
1035+564 at 1420 MHz band on five occasions. 
Table \ref{phasecal} gives the details of the phase 
calibrator 0834+555 at all other occasions and in all the frequency bands.

Tables \ref{1035cal} and \ref{phasecal}
give the positions of the phase calibrator obtained from GMRT
observations and their offset from the optical position of
the source taken from NASA Extragalactic Database (NED). 
We note that in earlier observations, the offset in the positions is
sometimes significant. This is because until June 2001, the GMRT
antennas were pointing at the apparent positions of the sources
(uncorrected for the effects due to nutation and aberration etc.)  and
was not corrected to their mean positions. Thus some of the observations gave
large position offsets, sometimes as large as few tens of arcseconds.
However, we found in the cases where the phase calibrators had large  offsets
from the optical positions, that, all sources in the
supernova field of view were offset from their respective
optical positions by the same amount.

\subsection{Data Analysis}


Data analysis at high frequencies is  easier as compared
to that for lower frequencies. At high frequencies, the bandwidth smearing
across the observing frequency band is negligible in contrast to the case for
low frequencies. Bandwidth smearing occurs due to the averaging of
visibilities over a finite bandwidth  and is directly
proportional to $\Delta \nu/\nu$. This is large at low frequencies
and leads to significant smearing. One needs to divide the total band
width of observation, $\Delta \nu$ into sub-bands $\delta {\nu}_i$ so
that $\delta {\nu}_i/\nu $ is small enough for insignificant smearing
and then stack the sub-bands together, in order  not to lose
the sensitivity.   At high frequencies, the w-term can be
ignored and a 2D approximation of the sky is valid.  At lower
frequencies however this assumption breaks down. 
This is because at low frequencies, the antenna primary beam
($\lambda/ D$, where $\lambda$ is the wavelength of observation and $D$
is the antenna diameter) is much larger to be approximated by a single
tangent plane.  Three-dimensional imaging is required which is
performed by dividing the whole field of view in multiple sub fields
and imaging each field separately. We find that for GMRT the minimum
number of sub fields required are 2, 6, 11 and 15 at 1420 MHz, 610 MHz, 
325 MHz and 243 MHz frequency bands, respectively
\footnote {Number of planes required at
a given wavelength $\lambda$ is $N_{planes}= \lambda B/ D^2$, where
$B$ is the longest baseline separation
(25 km for GMRT) and $D$ is the antenna diameter (45 m for GMRT)}.

Astrophysical Image Processing System (AIPS) developed by NRAO
was used to analyze all datasets with the standard GMRT data reduction.
Bad antennas 
and corrupted
data was removed using standard AIPS routines. Data were then
calibrated and images of the fields were formed by Fourier inversion
and CLEANing using AIPS task IMAGR (see
\citet{bha00,bha01} for details of data analysis procedure). 
For 1420 MHz and 610 MHz datasets, bandwidth smearing effects were
negligible and imaging was done after averaging 100 central frequency
channels (out of total 128 channels). For 325 MHz analysis, the
96 central channels (12 MHz bandwidth) were divided into 4 sub-bands of
3 MHz each for imaging. For 243 MHz analysis, we usually took 
60-70 good central channels and divided them in 4 sub-bands for imaging. While
making images, these sub-bands were stacked together. To take care of
the wide field imaging, we divided the whole field of view in 2-5 sub
fields for 1420 MHz frequency datasets, and in 5-9 sub fields in the 610
MHz datasets. For 325 MHz datasets, wide field imaging
was performed with 16-25 sub fields, while for 245 MHz datasets
this was done with 25-36 sub fields. 
We performed few rounds of phase self calibrations in
all the datasets to remove the phase variations due to the bad weather
and related causes and improved the images considerably.  After
imaging, all  the subfields were combined into a single field using  AIPS
task FLATN. All datasets were corrected for the primary beam pattern of the
antennas. Fig \ref{SN_FOV} shows the SN 1993J image on 2002 Mar 08 
at 243 MHz band.
The figure on the left hand panel shows the full field of view (FOV) which was
imaged after dividing it in to 25 subfields and subsequently
recombining. Positions 
of the supernova, the parent galaxy M81 and the nearby starburst
galaxy M82 are shown separately in the big panels.  Table \ref{tab:1} gives
the details of the observations and analysis of all the datasets
at GMRT observation epochs.

\subsection{Errors in the flux density determination}

Apart from the noise in the map plane (${\rm rms} \propto {[n(n-1)\Delta
\nu \,\tau]}^{-1/2}$ where $n$ is the number of baselines, $\Delta
\nu$ the bandwidth of the observations and $\tau$ the integration time),
there are other factors which contribute to the
errors in the flux density determinations. These could be modulations
due to interstellar scattering and scintillation (ISS), elevation
dependent errors, errors due to the bandwidth smearing and errors due to 
the use of 
two-dimensional geometry in imaging.  The last two errors can be
corrected for by following the procedure mentioned in the last
section. Here we discuss mainly the ISS effects and elevation
dependent errors which contribute to the uncertainty in the 
source flux density determination.

\subsubsection{Interstellar scattering and scintillation}

We calculated the effect of the interstellar scattering and
scintillation (ISS) in the SN at all observed frequencies.  Figure 1
of \citet{wal98} suggests that the transition frequency ${\nu}_0$ (the
frequency corresponding to unit scattering strength \footnote{
Scattering strength is unity when the phase change across the first
Fresnel zone introduced by ISM inhomogeneities is of the order of half
a radian \citep{wal98}}) for the 
position of SN 1993J ($l=142.15,\, b=40.92$) is 8 GHz.  Since all
the frequencies of our observations are less than the transition
frequency, the target should be in strong scattering regime.
Figure 3 of \citet{kul98} suggests that strong short-time diffractive 
scintillation occurs in very early stages of the supernova, when its
size is still very small. Their figure suggests that SN 1993J is in
the refractive interstellar scintillation (RISS) regime at the present
epoch. During the GMRT observations epoch, i.e. around day 3000 since
SN 1993J's explosion, we use the size of the SN to be $ {\theta}_s
\sim 5000\,\mu$as \citep{bar02}. Fresnel size of the source is
${\theta}_r \propto {({\nu}_0/\nu)}^{11/5}$ \citep{wal98}. We thus find
that the SN is in the point source limit only for 243 MHz 
since for this frequency, 
 ${\theta}_s \le {\theta}_r$ (${\theta}_r \sim 9000\,\mu$as) 
and is in the extended source limit (${\theta}_s \ge {\theta}_r$) for
all other frequencies ($ {\theta}_r \,{\rm is }\, 
\sim 4500\,\mu$as at 325 MHz, $\sim
1150\,\mu$as at 610 MHz, and $\sim 200\,\mu$ at 1420 MHz).
For RISS in the point source regime, the modulation index
$m_p$ (rms fractional flux variation) and the refractive scintillation
timescale $t_r$ are: $$m_p ={(\nu/{\nu}_0)}^{17/30}\,\,{\rm and}\,\,
t_r ({\rm hours}) \propto {({\nu}_0/\nu)}^{11/5}$$ respectively.  In
the extended source regime, these factors are further modified. 
The modulation index is reduced by ${({\theta}_r/{\theta}_s)}^{7/6}$
and refractive time scale is increased by
${\theta}_s/{\theta}_r$ \citep{wal98}.  
We calculate that at 243 MHz, the modulation index is
13.8\%, whereas time scale of modulation is $\sim 180$
days. The modulation index and the time scale of modulation at 325 MHz are
14.7\% and $\sim 100$ days respectively.  At 610 MHz band, the
modulation index is 4.2\% whereas the time scale of variation is 100
days. At 1420 MHz, the modulation index (0.8\%) is negligible and time
scale of variation is 100 days. We, therefore, conclude that the
interstellar scintillations and scattering are not likely to 
significantly modulate
our flux densities on short time scales (such as $\sim 2$ months).

\subsubsection{Elevation dependent errors}

At high frequencies, there are noticeable changes in the antenna gains
with change of elevation angle with time.  
Atmospheric opacity also introduces an
elevation dependence on the observed visibility amplitudes.  By
calibrating the target source with a nearby calibrator,  these
variations can be accounted for to a large extent.  
However, if the primary flux calibrator is
observed at a different elevation from the secondary gain and phase
calibrator, it may lead to significant errors.  Proper
calibration of the flux densities at high frequencies requires
the knowledge of a gain curve for the antennas, as well as the atmospheric
opacity. For low frequencies such as 610, 325 and
243 MHz, elevation dependent errors are quite negligible for most of
the sources. At 1420 MHz waveband, elevation dependent errors may
affect the observations. However, SN 1993J is at such a high
declination (69$\arcdeg$) that the data is unlikely to be
severely affected by
such errors. We ran AIPS task ELINT to determine the elevation
dependent errors for the phase calibrators. We find that the errors
are $\sim 5-6\%$ in 1420 MHz band and $\sim 2-4\%$ in 610, 325 and 243
MHz bands.  However the actual errors may be a little larger than this
because the program source (SN 1993J) is another 15-20$^o$ away from
the phase calibrators.

\subsubsection{Calibration errors}

In some of the observations, we had more than one flux calibrators.
In such instances we could estimate the calibration errors 
in GMRT bands by using
the following procedure.  We assumed one flux calibrator (say $ A$ ) to
be the absolute calibrator and calculated the flux density of the absolute
calibrator using Baars formulation. We then calibrated  another
flux calibrator (say $B$) with respect  to the absolute calibrator ($A$) and
obtained flux density of the calibrator $B$ with 
respect to $A$ (say $B(A)$). 
Since we already had the flux density of the flux calibrator 
$B$ using Baars formulation (say $B(Baars)$),
difference of $B(A)$ and $B(Baars)$ : $\mid B(A)-B(Baars)\mid$ is 
an indicator of the  calibration errors in GMRT. We found that
the calibration errors for 1420, 610, 325 and 243 MHz bands are in the
range 3-4\%, 4-6\%, 5-7\% and 5-8\% respectively.

To incorporate all the errors discussed above, we have used the
following formula to quote the errors in the flux density determination
for our GMRT datasets $$ \sqrt{[( {\rm map\, rms})^2 + (10\%\, {\rm
of\,the\,peak\,flux\,of\,SN})^2]}$$ The overall
error of 10\% of the peak flux density at all the frequencies
takes into account any possible systematic error in GMRT measurements
including the
small errors due to antennas' elevation dependence. This 
10\% is close to the actual 
errors in the low frequencies 240 and 325 MHz bands but is an upper
limit for the higher frequencies 1420 MHz and 610 MHz, where the
overall errors are more likely to be 7-8\%.

\section{Results and Interpretation}


\subsection{Light curves of SN 1993J}

In Fig. \ref{LC} we show the light curves of SN 1993J in 
all the frequency bands reported in Table \ref{tab:1}.
We interpret here 
the gross features of these light curves.  
The light curves in the 1420 and 610 MHz 
bands show  a decline with time.
This is the expected behavior of a supernova  in the 
optically thin part of the light curve, where the decline is
a power law in time. There appears to be a 
hint of a jump in the light curve somewhere between 3123 to 3296 days
in the 1420 MHz frequency band.
If the variation is real, the most natural explanation of this
variation could be the inhomogeneity in the 
circumstellar medium, or non-uniform magnetic field in the plasma.
\citet{bar02} also found a sudden jump in the 8.4 GHz light curve around 
day 2000. We  consider it likely that the jump in the 1420 MHz light curve
referred to here is real
and since the evolution of SN light curves repeat at lower frequencies at
later times, the earlier jump in the 
8.4 GHz light curve is being seen later in the 1.4 GHz band. 
However, in view of the uncertainties in the flux determinations, this
conclusion is not robust at present. 
Future observations at lower frequencies will 
be a good check of the possible jumps in the SN light curve, for 
if jumps are real then they will show up at lower frequencies at later 
epochs.  The 325 MHz light curve also shows a
declining trend, although in a less rapid manner compared
to the light curves at  higher frequencies. 
It is likely that the supernova
has just become optically thin in the  325 MHz band and hence the
decline is much slower. However, the 235 MHz 
light curve is more or less flat, as is expected when the light curve 
is going through a peak. 
Shape of the spectral fits in radio also suggests this (see next section)
and indicates that the light curve of the SN 1993J may remain flat
at 235 MHz band for the next few years.

Because of a small number of  data points, we do not attempt the
detailed fits of the light curves
with various models. However, we attempted to compare our data points with the 
already existing models. We used the model of synchrotron radio emission with 
free-free absorption \citep{wei02}. 
The free-free absorption model for the SN 1993J light curves
\citep{wei02} gives :
\begin{equation} \label{weiler1}
F_{\nu}({\rm mJy})=K_1 {\left(\frac{\nu}{5\,{\rm GHz}}\right)}^{\alpha} 
{\left(\frac{t-t_0} {1\,{\rm day}}\right)}^{\beta}\,e^{-\tau}
(1-e^{-{\tau}'}){{\tau}'}^{-1}
\end{equation}
where
\begin{equation} \label{weiler2}
\tau=K_2{\left(\frac{\nu}{5\,{\rm GHz}}\right)}^{-2.1} 
{\left(\frac{t-t_0} {1\,{\rm day}}\right)}^{\delta}
\end{equation}
and
\begin{equation} \label{weiler3}
{\tau}'=K_3{\left(\frac{\nu}{5\,{\rm GHz}}\right)}^{-2.1} 
{\left(\frac{t-t_0} {1\,{\rm day}}\right)}^{{\delta}'}
\end{equation}
Here $K_1$, $K_2$, and $K_3$ are free parameters corresponding to the
flux density, uniform and non-uniform external absorption
respectively at 5 GHz, one day  after the explosion date $t_0$. Free
parameters $\alpha$ and $\beta$ are the emission index and time index
respectively.  The parameters $\delta$, ${\delta}'$ describe the time
dependence of the optical depths for the uniform and nonuniform
circumstellar media.  The best fit parameter values for SN 1993J 
\citep{wei02} are:
$K_1=1.86 \times 10^4$, $\alpha=-1.07$, $\beta=-0.93$,
$K_2=1.45 \times 10^3$, $\delta=-2.02$, $K_3=6.31 \times 10^4$,
and ${\delta}'=-2.14$.  This model was fitted for high frequency VLA
observations in the
range 22.5 GHz to 1.4 GHz frequency bands.
 We extrapolate this model to 1420, 610, 325 and
243 MHz frequencies at GMRT observation epochs with the above 
parameters. Fig \ref{lc.wei} shows this extrapolated light
curves for the SN at 1420, 610, 325 and 243 MHz and our corresponding GMRT
data points at the respective frequencies. It is evident that the
free-free model described above overpredicts the flux densities at low
frequencies. In fact lower the frequencies, more significant is the
departure from the standard free-free model. This indicates that 
the optical depths fitted using high frequency datasets, 
simply extrapolated to low frequencies with the
dependence $\tau \propto {\nu}^{-2.1}$ are not sufficient to account for the 
required absorption.
 One needs to incorporate some additional frequency dependent 
opacity at low frequencies, which can compensate for the difference
between the model light curves and the actual data. 

Since the extrapolated free-free model of \citet{wei02}
with  the above parameters to low frequencies, say 1420, 610, 325 and 243
MHz tends to overpredict the flux density of the supernova, we
incorporated an extra opacity due to synchrotron self absorption 
to check whether this brings the observed flux points into consistency. 
The SSA absorption
coefficient is given as \citep{pac70}
\begin{equation}
{\kappa}_{SSA}=\frac{\sqrt{3} e^3}{8\pi m}
{\left(\frac{3 e}{2\pi m^3 c^5}\right)}^{\gamma/2}c B^{(\gamma+2)/2}
\Gamma\left(\frac{3 \gamma +2}{12}\right)
\Gamma\left(\frac{3\gamma+22}{12}\right) 
{\nu}^{-(\gamma+4)/2}
\end{equation}
We assumed shell size (from where radio emission is coming)  
to be 20\% \citep{bar02} of the supernova size and used SSA optical depth as
${\tau}_{SSA}={\kappa}_{SSA} R/5$ with $R \propto t^{0.781}$ \citep{bar02}.
Eq. \ref{weiler3} is the internal absorption term, to which we add the
SSA optical depth ${\tau}_{SSA}$ i.e.
\begin{equation} \label{with_ssa1}
\tau_{int}=  {\tau}'+{\tau}_{SSA}
\end{equation}
And the flux density of the SN including the SSA absorption is 
\begin{equation} \label{with_ssa}
{F_{\nu}}^{SSA}({\rm mJy})=K_1 {\left(\frac{\nu}{5\,{\rm GHz}}\right)}^{\alpha}
{\left(\frac{t-t_0} {1\,{\rm day}}\right)}^{\beta}\,
\left(\frac{1-e^{-{\tau}_{int}}}{{\tau}_{int}}\right)
\end{equation}

\noindent where the bracketed term is the attenuation due to the total internal
absorption including SSA absorption.
Fig. \ref{lc.wei} shows the light curves at 1420, 610, 325 and 
243 MHz after incorporating
the extra SSA opacity. We note that 
even after adding this extra opacity in the existing free-free model, 
the total optical depth is not sufficient to 
reproduce the observed flux density.
The medium is more opaque to the radio emission at the lower frequencies.


\subsection{Spectra of SN 1993J}

On five different
occasions we collected simultaneous or near simultaneous spectra of SN1993J
by making flux density measurements at multiple frequencies separated
by about a month or so. Since the supernova is
$\sim$ 10 years old, we do not expect its flux density to change by
significant amount in a month's gap between the two 
flux density measurements. For
example, using the flux density dependence on time $t$ as $F\propto t^{-0.93}$
\citep{wei02}, we find that the flux density 
of the supernova will change only by 2\% in 75 days at such late epochs.
This is well within the uncertainties of the flux measurements.
We also saw in the last section that the time scale of the flux density
variation due to RISS is 6 months (3 months) at lower (higher)
frequencies and hence RISS will not contribute significantly to 
the flux density
variations within a month's time gap.  
Therefore, we can use the observations separated by
roughly a month to obtain a near-simultaneous spectrum of the SN.  Table
\ref{tab:2} gives the details of the observations used to obtain the
spectra with GMRT at five different epochs.

                                             
With so few data points in the spectra it is difficult to pinpoint the 
synchrotron self absorption model from the free-free absorption mechanism as
the dominant underlying cause of absorption.
We use synchrotron self absorbed radio emission model to fit our datasets
(See \citet{che98,pac70}), which we justify in the next section.
\begin{equation} 
{f_{\nu}}^{SSA}=\frac{\pi R^2}{D^2} \frac{c_5}{c_6}B^{-1/2}{\left(\frac{\nu}{2c_1}\right)}^{5/2}[1-g(\nu,\gamma)]
\end{equation}
where
\begin{equation}
g(\nu,\gamma)={\rm exp}\left[-{\left(\frac{\nu}{2c_1}\right)}^{-\frac{\gamma+4}{2}} \left(\frac{4 f R c_6 N_0 B^{\frac{\gamma+2}{2}}}{3}\right)\right]
\end{equation}
 and
\begin{equation}
N_0=\frac{a B^2 (\gamma-2) {E_l}^{\gamma-2}} {8 \pi}
\end{equation}
The above flux density expression can be reduced to the following in
the optically thin limit.
\begin{equation}\label{thin}
{{f_{\nu}}^{SSA}|}_{\tau\le1}=\frac{4\pi f R^3}{3 D^2} c_5 N_0 B^{(\gamma+1)/2}{\left(\frac{\nu}{2c_1}\right)}^{-(\gamma-1)/2}
\end{equation}
And in the optically thick limit
\begin{equation}\label{thick}
{{f_{\nu}}^{SSA}|}_{\tau\ge1}= \frac{\pi R^2}{D^2} \frac {c_5}{c_6}B^{-1/2}{\left(\frac{\nu}{2c_1}\right)}^{5/2}
\end{equation}
Here $R$ is the radius of the supernova, $D $ is the distance to the
supernova, $f $ is the filling factor (which we took to be 0.5 for our
purpose  \citep{che98}); $\gamma$ is the electron spectral index
which depends on the emission spectral index $\alpha$ as $\gamma=2 \alpha
+1$.  $N_0$ is the normalization constant in the power-law distribution of
electrons in $N(E)=N_0 E^{-\gamma}$ \citep{che98}, and $a$ is the
equipartition factor ($a=U_{rel}/U_B$; $U_{rel}$ is the relativistic
electron energy density, and $U_B$ is the magnetic energy
density), and $E_l$ is the electron rest mass energy ($E_l=0.51$ MeV). The
(constant) parameters 
$c_1, c_5, c_6$ are defined in \citet{pac70}; $c_5$, $c_6$ are
functions of electron spectral index $\gamma$ and are tabulated in Table 7
of Appendix 2 in \citet{pac70}. They are:
 
$$c_1 = {3e \over 4\pi m^3 c^5}= 6.27 \times 10^{18} \; \rm c.g.s.\; \rm units$$

$$c_5 = {\sqrt 3 e^3 \over 16 \pi} 
{e^3 \over m c^2} \left( {\gamma + 7/3 \over \gamma +1}\right) \Gamma \left({3\gamma -1 \over 12}\right) \Gamma \left( {3\gamma + 7 \over 12}\right)$$ 

$$c_6 = {\sqrt 3 \pi  \over 72} e m^5 c^{10} \left( \gamma + {10\over3} \right) \Gamma \left({3\gamma +2 \over 12}\right) \Gamma \left( {3\gamma + 10 \over 12}\right)$$ 
%
The parameters
$R$, $B$ and $\alpha$ or $\gamma$ are the three free parameters for the 
fitting a model to the observed spectrum. 

On day 3200, we combined the GMRT low frequency spectrum  with that
of high frequency VLA data (kindly provided by
Kurt Weiler, C. Stockdale and their collaboration), 
and fit the synchrotron self absorption (SSA) model (solid line in
Fig \ref{comb}). 
We found that the spectrum at high frequency is best fit by a broken power
law with a break around 4 GHz which we interpret as due to synchrotron cooling.
SN 1993J is the first such young supernova where the synchrotron cooling break 
in the radio frequency region is seen.
Observational signature of the synchrotron cooling break directly leads to the
determination of the magnetic field in the plasma. This
is the most direct determination of magnetic field in the shocked plasma
\citep{cha04}, since the age of the source is known.
The lifetime of the relativistic electrons undergoing synchrotron loss is
given as
\begin{equation}
t=\tau = E/[-{(dE/dt)}_{Sync}]=1.43 \times 10^{12} B^{-3/2} {\nu}^{-1/2}_{break}\; {\rm sec}
\end{equation}
Here we use  ${B_{\perp}}^2={(B\, {\rm Sin \theta})}^2 =(2/3)B^2$.
However, we also include energy loss/gain due to the adiabatic expansion
and the diffusive shock acceleration (Fermi acceleration)
along with the synchrotron energy 
loss term, for the SN is young and these process 
are likely to be important (see 
\citet{cha04}). Hence, the lifetime of electrons  
for the cumulative energy loss rate is
\begin{equation}
\tau = \frac{E}{{(dE/dt)}_{Total}}=\frac{E}{ (R^2 t^{-2}/20 {\kappa}_{\perp})E
-b B^2  E^2- t^{-1}E}
\end{equation}
where the second term in the denominator is synchrotron loss term with
$b=1.58 \times 10^{-3}$; the 
first term is the energy gain due to diffusive shock 
acceleration and the third term is the adiabatic expansion loss term.
Using ${\nu}_{break}=5.12\times 10^{18} B
 E_{break}^2 \;{\rm Hz}$ \citep{pac70} in the above equation, we found the 
magnetic field implied by the life time argument 
to be $B=330$ mG. On the other hand, from the best fit with SSA, the magnetic 
field under equipartition assumption is $B_{eq}=38\pm17$ mG.
Comparison of the magnetic field  obtained from the lifetime argument 
with that obtained from 
SSA model best fit  (assuming equipartition) determines the
value of the equipartition fraction between relativistic energy of
particles and magnetic field energy. Equipartition fraction
$a=U_{rel}/U_{mag}$  varies with magnetic field $B$ as
$a = (B/B_{eq})^{-(2 \gamma +13)/4}$ \citep{che98}. 
Therefore, the fraction $a$ ranges between $ 8.5 \times
10^{-6}\,-\, 4.0 \times 10^{-4}$ with a central value 
of $a=1.0 \times 10^{-4}$ (corresponding
to $B_{eq}=38$ mG) on day 3200.
As \citet{che98} has argued, it is likely that this 
high magnetic field is the result of an amplification process in
the interaction region and is not merely the effect of shock compression
of the magnetic field in the wind by the relevant compression ratio. 
See \citet{cha04} for a detailed study of the combined spectrum
(GMRT and VLA) with the break. 

The above analysis shows that if we were to fit the SSA model under
the assumption of equipartition, we would end up seriously underpredicting the
magnetic field, roughly an order of magnitude lower than the
actual field. The relativistic plasma is far from equipartition
and is strongly dominated by the magnetic energy density. 
We assume that the equipartition fraction, as determined above on 
day 3200 does not change with time. Using this fraction, we can 
fit the synchrotron self absorption  model to our subsequent
GMRT spectra and determine the most probable values of the 
free parameters, e.g. $R$, $B$, and $\alpha$ in the SSA model. 
Since we
have very few data points for the spectra, we cannot determine all  three
parameters. Therefore we separately calculated the value of spectral
index $\alpha$ using the optically thin part of the spectrum ($F_{\nu}
\propto {\nu}^{-\alpha}$) between frequencies 610 MHz to 1420 MHz. The
errors in the values of the spectral indices reflect the errors in the
flux densities at a  given frequency. The size of the supernova and the
magnetic field were used thereafter as free parameters in the SSA fits to the
spectra. The best fit
magnetic field and the size of the supernova thus obtained and reported in 
Table \ref{bestfit} {\it are not under the assumption of equipartition}
of relativistic particles and fields. 
Errors in $B$ and $R$ are large due to a large range in the
equipartition fraction $a$ ($B \propto a^{-4/(2 \gamma +13)}\,; R
\propto a^{-1/(2 \gamma +13)}$).  
Fig. \ref{spectra1} shows the GMRT spectra at days 3000, 
3200, 3266, 3460 and 3730 since explosion. The spectrum currently appears to 
peak around 235 MHz due to the flattening of the light curves
at these frequencies and is gradually shifting to lower frequencies. 
Data in the still lower frequency  bands ($< 235$ MHz) may 
constrain the peak more tightly. 
We estimate below some of
the physical parameters of the supernova based on the spectral fits, 
and compare them with those determined by independent methods.

\subsection{Role of synchrotron self absorption at spectral turn-over}
We fit only the synchrotron self absorption model to the obtained spectra.
However, we have also tried to fit the free-free absorption model 
to the spectra. 
Since there are few data points in each spectrum, these models are not
easily distinguished against each other by the data. 
However the SSA model can be indirectly tested by comparing its
predicted parameters against similar quantities measured entirely 
independently. For example, SN 1993J has been 
extensively studied using VLBI and therefore the 
size of the SN is known from these measurements. We also obtain the
size of the SN from the SSA model using the 
turn-over in the spectra. Comparing the sizes 
obtained from these two independent methods therefore tests the SSA model. 
GMRT observations are only around  day 3000 and later.
To evaluate the SSA model
at earlier epochs, we used VLA data available on Internet
\footnote {URL: http://rsd-www.nrl.navy.mil/7213/weiler/kwdata/93jdata.asc}
starting from day 65 to day 250. We  digitized the VLA data for
day $400-800$ from the 
available light curves \citep{van94,fra98}.
We also used the published data of \citet{bar02} and \citet{per02}. We
fit the SSA model to the spectra at all these epochs using the
equipartition fraction derived in \citet{cha04} and obtained the best
fit magnetic field and size of the supernova. We compared the best fit
size so obtained with that of VLBI size of SN at various epochs
obtained from Fig 6 of \citet{bar02}. Fig \ref{size} shows the SSA
size of the supernova plotted against the VLBI size of the
supernova. At late epochs beyond day 3000, 
no VLBI observations were available, so we
have extrapolated the earlier VLBI observations to obtain the sizes
of the SN at the relevant epochs, assuming the latest value of $m=0.781$
in $R \propto t^m$ \citep{bar02}. We note that the VLBI size of the
supernova and the SSA model fit size from the peak of the spectrum are
largely consistent at all epochs. If SSA  was not 
the most important absorption mechanism, the SN size determinations
from the SSA model would have been smaller than the sizes obtained from 
VLBI (see \citet{sly90}).
Since the two are roughly consistent at all epochs, the
conclusion that synchrotron self absorption is the dominant absorption 
mechanism to determine the spectral turn-over appears natural.

We notice that at late epochs
beyond day 3400, the SSA radius is
more than the extrapolated VLBI size. 
This could be due to the
incorrect estimation of the  
extrapolated VLBI radius with the assumed $m$,
at late epochs for which VLBI observations do not exist.
Additionally,
uncertainties in the determination of the peak of the spectra 
due to the lack of very low frequency measurements may
affect the accuracy of the SSA radius and associated parameters. 

\subsection{Evolution of spectral index, size and magnetic field}

Fig \ref{alpha.mag} shows the evolution of the best fit radio 
spectral index $\alpha$. The spectral
index 
evolves with time and its value changes from
$\alpha \sim 0.8-1.0$ (in the first few tens of days) to 
$\sim 0.6$ (10 years
after explosion). GMRT measurements of the spectral index 
have significant  errors.
This is because we have not fit the spectral index
with SSA or FFA model since we had very few frequency data points in
the spectra. Instead we calculated spectral index from the optically
thin part of the GMRT spectra between frequencies 610 and 1420 MHz
frequency bands, using the relation $F_{\nu} \propto
{\nu}^{-\alpha}$. Errors in the determination of $\alpha$ are partly due to
the errors in the flux density measurements.

Fig \ref{size.mag} (upper panel) shows the supernova size 
evolution with time. We also plot the indicative $R \propto t$ line
in the Figure. It is noted that at early epochs, the size evolution
is consistent with $R \propto t$, i.e. supernova ejecta expansion is
free expansion while at late enough epochs, the expansion 
undergoes deceleration.
 These results are roughly consistent with those of \citet{bar02}
and \citet{mar97}. We show the time variation of the radii determined 
from our SSA fits to GMRT spectra as an inset in Fig \ref{size.mag}. 
The lower panel of Fig \ref{size.mag} shows that the
magnetic field decreases as supernova ages. 
We also
plot the indicative $B \propto t^{-1}$ line, and it
suggests that within  error bars, the magnetic
field decreases according to $B \propto t^{-1}$ on the full time range.  
However there 
seems to be a hint of flattening in the time development
between  $\sim 400-1300$ days.  

When synchrotron self absorption is the dominant absorption mechanism,
it is not possible to estimate directly the circumstellar density, as
is possible in the case of {\it dominant} external free-free absorption;
but in contrast it is possible to estimate the radio supernova size, 
if the radio peak flux is observed.
This leads to the determination of the velocities of the outer parts of the
supernova, which we find from the radii reported in Table {\ref{bestfit}
to be: $v \approx 11000\,  { \rm km\,s^{-1}}$ for $t \geq 3000 \; \rm day$. 
As a comparison, note that the optical photosphere velocity
decreased to $7000-10,000 \; \rm \; km \; s^{-1}$ range even $10 - 30 \;
\rm day$ after explosion (\citet{ray93}).

\subsection{Mass loss rate of the progenitor star}

Estimates of the mass loss rate give crucial information about the
progenitor star. Since typical ejecta velocities are $\sim 1000$ times
that of wind velocities, information about mass loss rate derived a few years
after the explosion can trace the history of the progenitor star a few
thousands of years before the explosion. 
This requires an assumption about the
wind velocity for the mass losing progenitor 
star when it exploded as a supernova.  The wind velocities around red
supergiant stars are $\sim 10$ km s$^{-1}$ while that of blue
supergiant stars are $\sim 1000$ km s$^{-1}$. For SN 1993J, we assume
in consonance with data and models (see e.g. \citet{ray93})
that the progenitor star had an
extended red supergiant like envelope in its last stage of evolution;
thus we adopt the circumstellar wind velocity to be $ 10$ km s$^{-1}$.
We then 
determine the mass loss rate using the best fit
parameters of SSA to the GMRT data (Table \ref{bestfit}).

It is clear from section 3.2 that the magnetic energy density in SN 1993J 
considerably exceeds the relativistic particle energy density. However,
these ratios are close to unity in
some other  supernovae, e.g. SN 2002ap \citep{bjo04}, SN 1998bw \citep{kul98}
etc. \citet{fra98} have also shown that there is a rough equipartition
between the magnetic energy density and the thermal energy density
in the post-shock gas in SN 1993J. In turn these are related to 
the ram pressure of the shock front.
This leads to an estimate of the mass-loss rate from
the SN progenitor \citep{che98} by relating the post-shock magnetic energy 
density to the shock ram pressure: $B^2/8 \pi = \zeta \rho_0 v_{sh}^2$, 
where $\zeta$ is a numerical constant, $\rho_0$ is the post-shock density 
and $v_{sh}$ is the shock velocity. 
For $\zeta \le 1$, this gives a lower limit of the
estimate of the mass loss rate as
\begin{equation}
\label{massloss}
\dot M =\frac{ 6.6 \times 10^{-5}}{m^2 \, \zeta }\, 
{\left[\frac{B}{0.33\,  \rm G}\right]}^{2}
{\left[\frac{t}{3200 \,{\rm days}}\right]}^{2}
{\left[\frac{v_w}{10\, {\rm km\, s^{-1}}}\right]}\,
{\rm {M}_{\odot}\, {yr}^{-1}}
\end{equation}

\noindent Here, $m=(n-3)/(n-2)$, $n$ being ejecta density power law, is the 
deceleration index.
Although there are no VLBI measurements for the epoch of GMRT
observations, we use the 
extrapolated value of the latest $m$ obtained from VLBI i.e. $m=0.781$
\citep{bar02}. We use the magnetic fields in the above equation
that are determined from the spectral fits as given in Table \ref{bestfit}.
For $\zeta =1$, the lower limit on the mass loss rate can be
calculated from Eq. \ref{massloss}. The values of the mass loss rate
for GMRT observations thus obtained are reported in Table \ref{mass_loss}. 
Fig \ref{mass} shows that the mass loss
rate  (obtained from GMRT observations)
appears to be decreasing slowly with a small slope of 
$1.8 \times 10^{-8} {\rm {M}_{\odot}\, {yr}^{-1}/yr}$.  (We assumed
a constant ratio of the ejecta velocity and the wind velocity to be
$v_{ej}/v_{w} \sim 1000$  to obtain the time before explosion 
from the epoch after explosion). Note that the predicted mass
loss rate in Chevalier (1998) is constant, since it goes as $B^2 t^2$
while his assumed time dependence is: $B \propto 1/t$. However,
GMRT data (from day $~3000$ to $3730$, -- a relatively short range
compared to the entire data span from the date of explosion) 
appears to indicate a slightly steeper dependence of
B on $1/t$. Although the best fit line to the mass loss rate
in Fig. \ref{mass} in the narrow range of time before explosion
shows a small decrement,
within the error bars of the GMRT measurements of B as seen 
Fig \ref{size.mag}, and on the longer
timescale since explosion
our results are consistent with a constant mass loss rate.

\section{Summary and Conclusions}

Even though SN 1993J has been one of the most well-observed
targets since its explosion some 11 years ago, it continues to illuminate the 
physics of supernovae. SN 1993J is a unique 
supernova for which magnetic field and sizes are determined from model
independent measurements; the former from the synchrotron cooling
break and the latter from VLBI measurements.
These have been utilized here to test models and parameters that determine
the radio emission from this SN.

Because the radio emission from SN 1993J
now peaks at frequencies lower than 235
MHz, it is necessary to observe it 
at lower frequencies, where the SN is 
still in the optically thick regime to properly determine the 
turn-over in the spectrum.
Multifrequency spectra of SN 1993J ranging from the very low GMRT frequencies 
to very high VLA frequencies will thus be important observational
inputs in future.
The synchrotron cooling break in the combined (VLA and GMRT) 
spectrum on day 3200 
leads to the determination of the equipartition fraction between 
relativistic particle and magnetic energy densities \citep{cha04}. 
From the self consistent analysis of early spectra
of SN 1993J, \citet{fra98} had interpreted the rough 
equipartition between the nonthermal ions, the magnetic field, and the
thermal energy.
We affirm that the 
relativistic particle energy density however,
is minuscule compared to the magnetic 
energy density in the radiating plasma. 
        
The outer part of the supernova from where the radio emission originates
is expanding at a speed of $\sim 11000 \rm \; km \; s^{-1}$ for 
$t \geq 3000 \rm days$. This speed is small compared to those found
in type Ic SNe like SN 1998bw, SN 2002ap or SN 2003dh (\citet{ray03})
indicating that ordinary type IIb supernovae like SN 1993J have  
much less extreme properties than other core collapse supernovae 
some of which may produce gamma-ray bursts.
Estimates of the mass lost from the progenitor star of SN 1993J derived
from post-shock magnetic pressure and shock ram pressure
and the GMRT spectra indicate that the mass loss rate remained roughly
constant during 8000-10,000 years before explosion.

In our earlier paper \citep{cha03}, we had argued that the size of
SN 1993J determined from the SSA fits
are roughly half of that obtained
from VLBI measurements. However, our earlier estimates of the sizes were based 
on the assumption of equipartition between magnetic 
and relativistic particle energy densities since we did not have a
direct handle on the equipartition fraction prior to the discovery
of the synchrotron cooling break.
In this paper we use the equipartition fraction directly determined 
from the synchrotron cooling break in the GMRT plus VLA spectrum on day 3200
\citep{cha04} and thereby obtain the best fit size of the SN.
We now find  that our derived SSA sizes 
roughly match the VLBI sizes of the supernova at all 
epochs. We thus affirm that even if free-free absorption may
have had a significant contribution to the total absorption in the
radio band, the peak in the spectrum is primarily determined by synchrotron
self absorption. \citet{sly90} also had plotted the SSA sizes of 
a few supernovae against their VLBI sizes and found
them to be consistent at all epochs.
He thence 
argued that SSA alone is responsible for the turn-over in the 
spectra of supernovae. However his results were based on the 
equipartition assumption between relativistic energy density and
magnetic energy density. Our results on SN 1993J confirm the conclusion of
\citet{sly90}, namely SSA is the main absorptive process
for the turn-over, but unlike 
\citet{sly90} we do not assume an equipartition. 
Light curves based on high frequency FFA models
extrapolated to low frequencies overpredict the flux densities
at low frequencies. Some extra opacity is needed to incorporate the
difference. We added an extra opacity due to synchrotron self absorption
which also could not fully account for the required absorption.
This suggests that the low frequency opacity in SN 1993J is not a simple 
extrapolation of high frequency opacity and a hitherto 
unaccounted for absorption may be at work at low frequencies.
%
%
%
%

\citet{bar02} have argued that the deceleration factor $m$ ($R \propto t^m$)
measured with VLBI 
dropped from 0.92 to 0.78 around day 400 and then around day 1500,
the decline of $m$ stopped and it increased to a value of 0.86. This trend
was roughly mirrored in the light curves. They argue that the upturn in 
$m$ could be due to the SN ejecta hitting the reverse shock and exerting 
pressure in the forward direction. A non-uniform 
radial dependence of the  circumstellar 
medium density can simultaneously provide a reasonable explanation of a
non-smooth evolution of the 1420 MHz flux between day 
$3150-3300$. 
The flux density may increase if the supernova ejecta hits
a clump in the CSM and decrease if the ejecta runs into a rarefied
region. The enhanced emission due to the 
interaction with a clump may in effect
lead to a flattening of the light curve rather than an actual
rise, if the filling factor of the clumps is low 
near the radiosphere. A non-smooth evolution of the radio luminosity 
can also be caused by inhomogeneities in the magnetic field in the
interaction region,
since synchrotron emission efficiency depends on the strength of
the magnetic field. 
The synchrotron flux density has a dependence on magnetic field as
${{f_{\nu}}^{SSA}|}_{\tau\ge1} \propto B^{-1/2}$ in the optically
thick part of the spectrum, and
${{f_{\nu}}^{SSA}|}_{\tau\le1} \propto B^{(\gamma+1)/2}$ in the optically
thin part. Therefore,
the variation in the flux density with magnetic field is less
significant in the optically thick regime than in the 
optically thin regime.
Hence, jumps in the flux densities at high frequencies 
will be more evident than at the lower frequencies. 
Due to the large errors of the measurements, one cannot
ascertain whether the sudden jump in the light curve at 1420 MHz is real 
(though the same had been seen in 8.4 GHz light curves at earlier epochs by 
\citet{bar02}).
But if they are real, then 
they should be mirrored in low frequency light curves at later epochs. 
Further observations at low frequencies will confirm it.

\acknowledgments
We thank Kurt Weiler for kindly providing the high frequency flux densities
from VLA observations on January 13, 2002. 
We thank the staff of the GMRT that is
run by the National Center for Radio Astrophysics of Tata Institute 
of Fundamental Research. We acknowledge the use of the 
Astrophysical Image Processing System (AIPS) which was developed
by the staff of National Radio Astronomical Observatory. 
We thank the anonymous referee for his/her detailed comments which helped
us to improve the presentation of this work. Poonam Chandra is a receipient
of the Sarojini Damodaran International Fellowship.
This research is part of the Tenth
Five Year Plan Project 10P-201 at TIFR, Mumbai.

\clearpage


\clearpage
\begin{figure}
\plotone{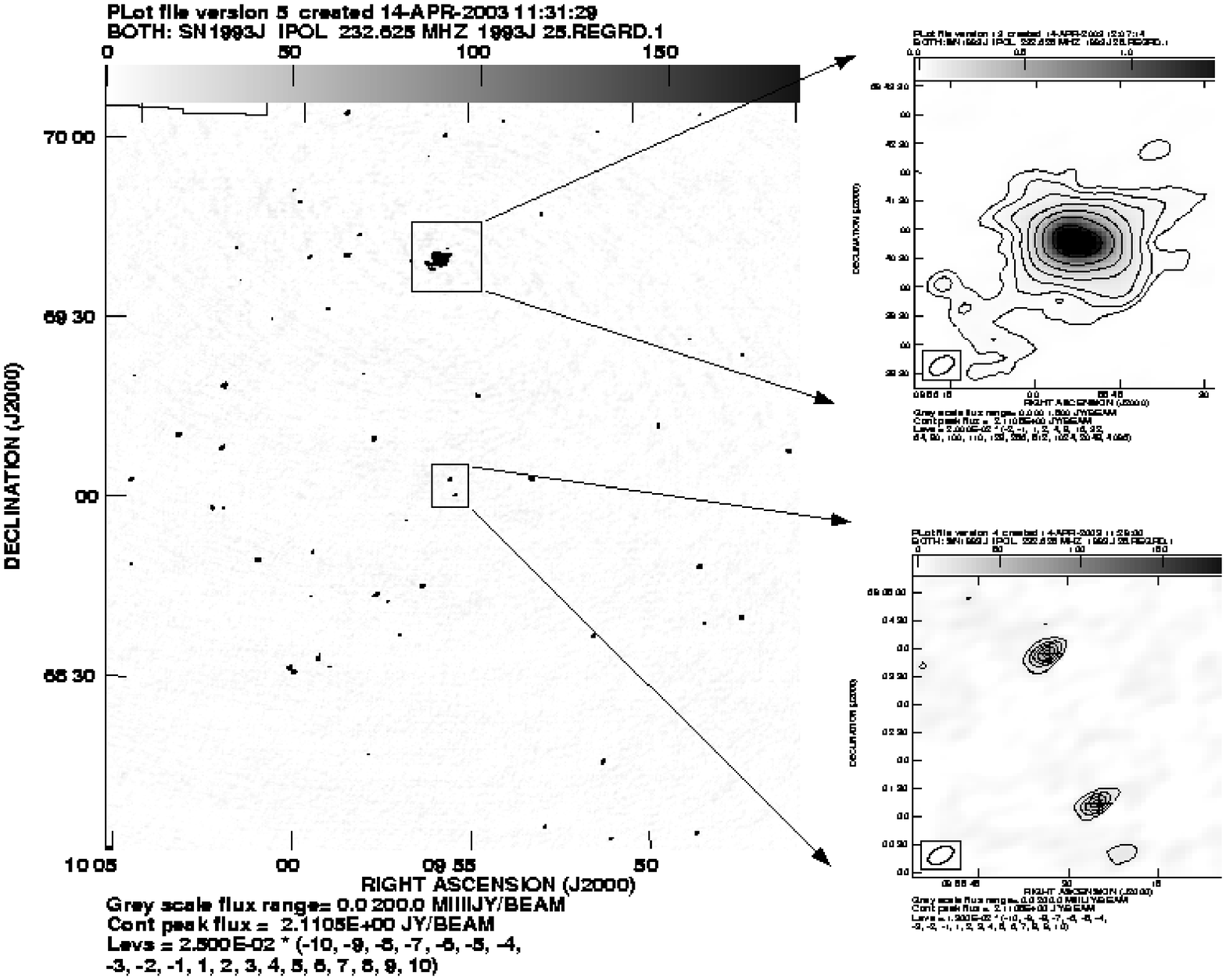}
\caption{GMRT contour map of the field of view
of SN 1993J in 243 MHz band (Observed on 2002 Mar 8). The upper square panel
contains M82 and lower square panel contains the parent galactic center
 M81 (Top) and SN 1993J (right).}
 \label{SN_FOV}
\end{figure}

\clearpage

\begin{figure}
\plotone{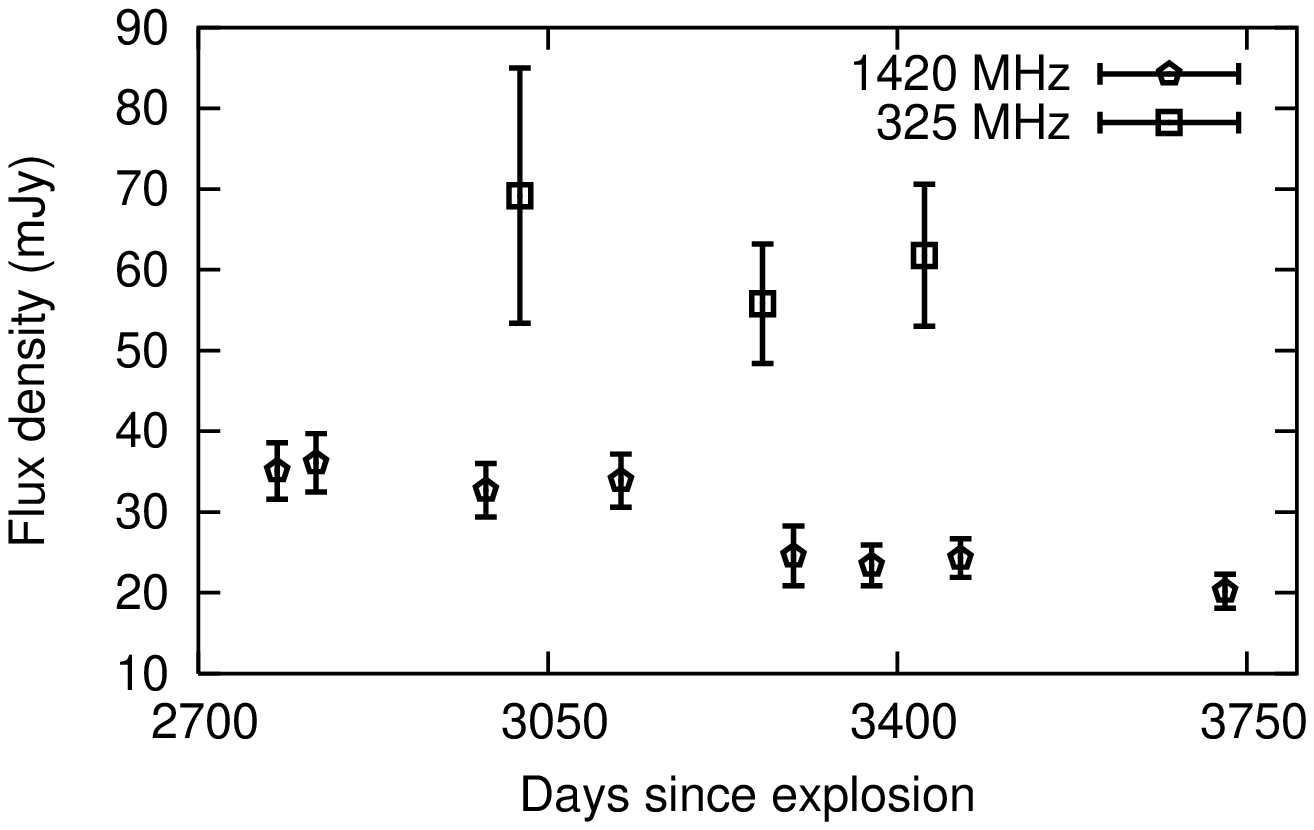}
\plotone{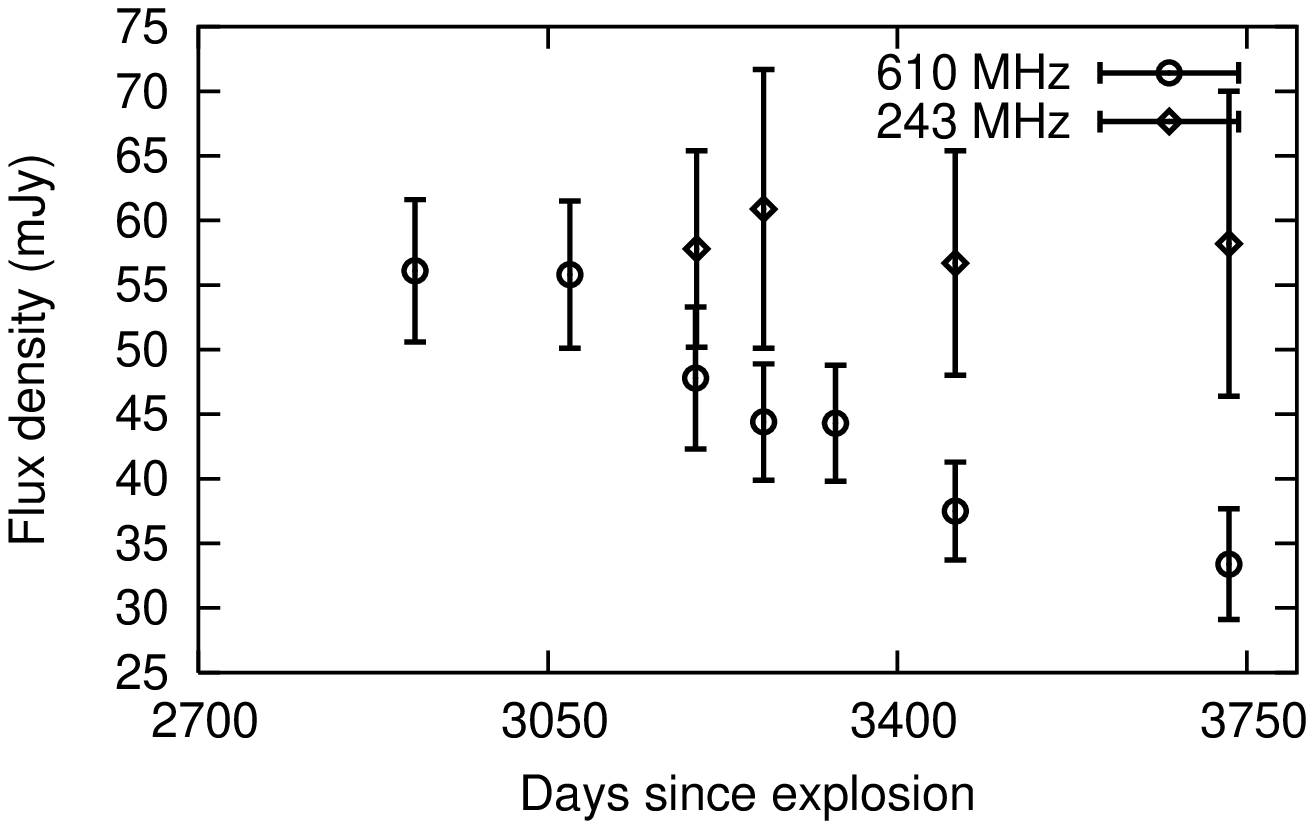}
\caption{Light curves of SN 1993J at 1420 MHz, 
325 MHz bands (upper panel) and 610 MHz and 235 MHz bands 
(lower panel) observed with the GMRT.}
\label{LC}
\end{figure}

%

\clearpage

\begin{figure}
\epsscale{0.90}
\plotone{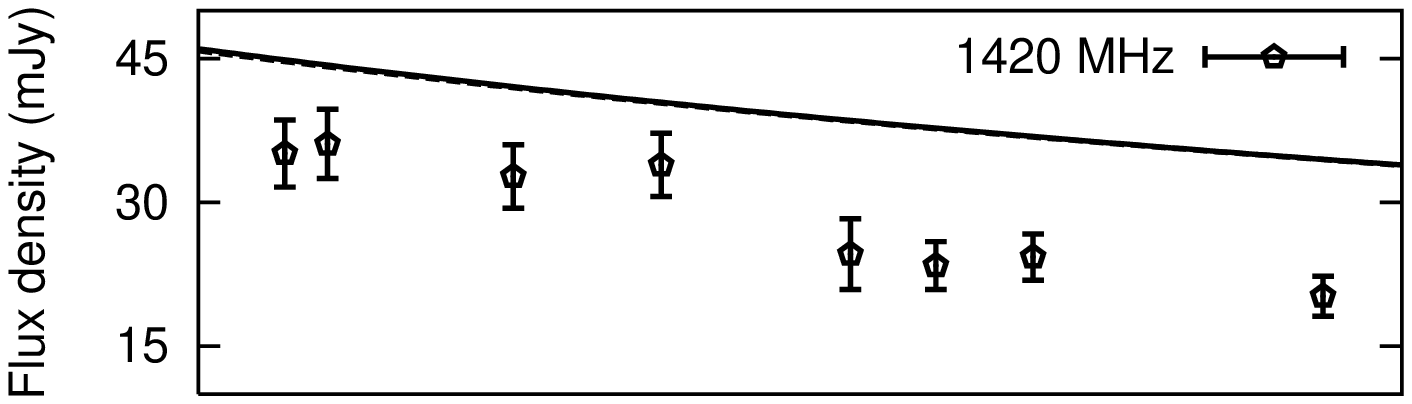}
\plotone{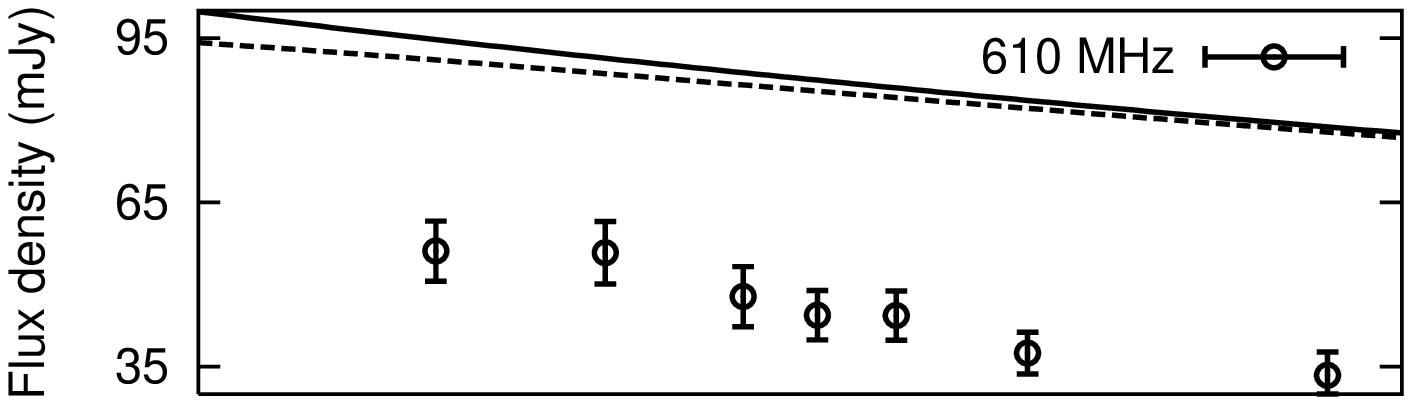}
\plotone{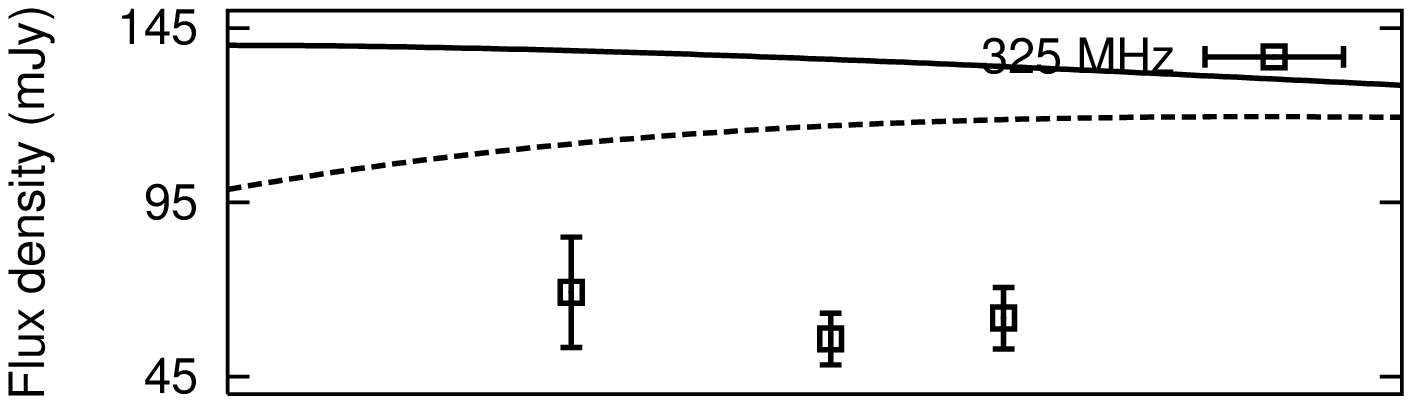}
\plotone{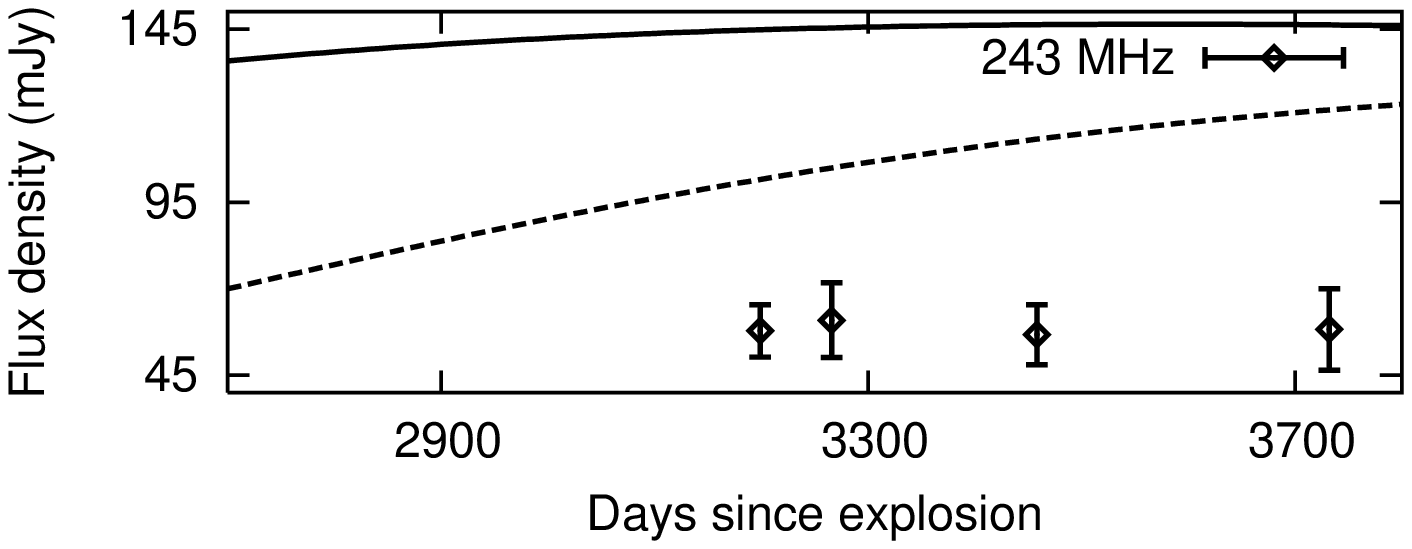}
\caption{Comparison of low frequency 
data to the predictions of models obtained by fitting
high frequency fluxes of SN 1993J. The  solid lines in
all four plots are Weiler et al's (2002)  model extrapolated
to lower frequencies (see sec. 3.1). 
Dashed lines are the flux density plots after 
incorporating the SSA optical depth in the Weiler et al. free-free model.} 
\label{lc.wei}
\end{figure}

\clearpage

%

\begin{figure}
\epsscale{0.40}
  \plotone{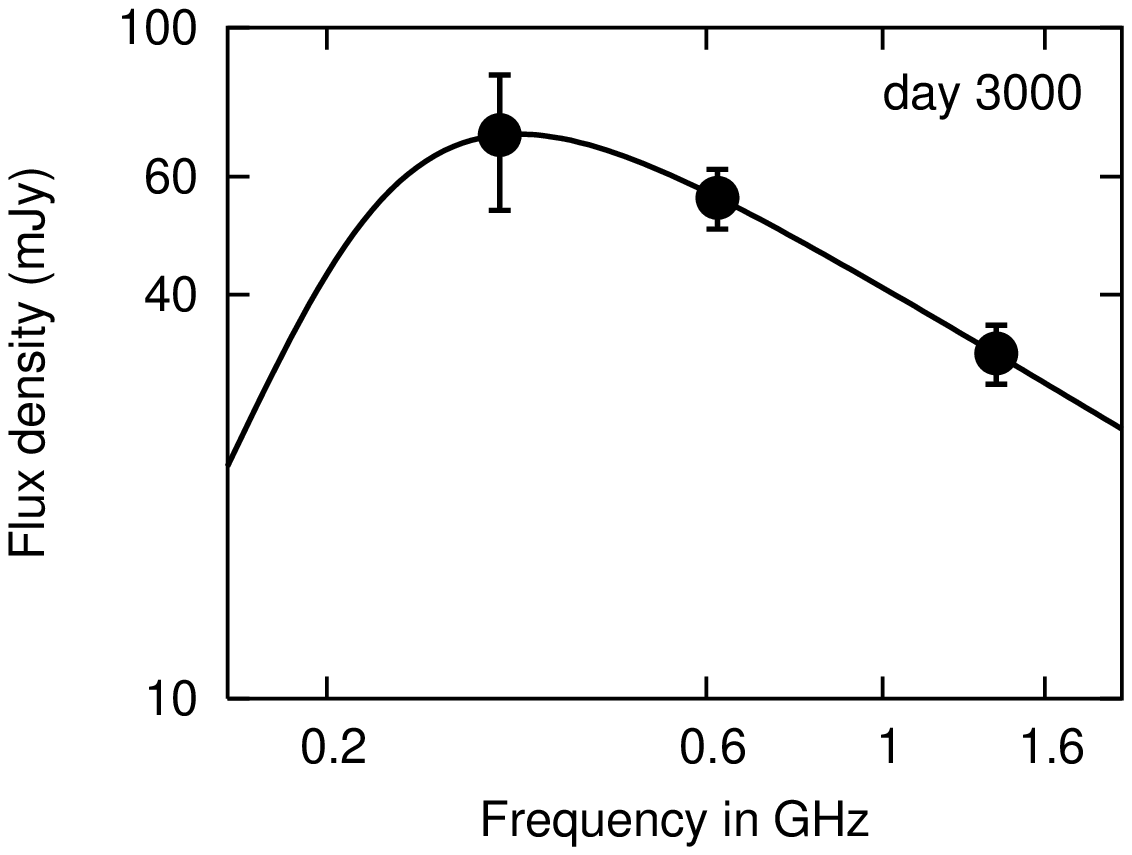}\hfil\plotone{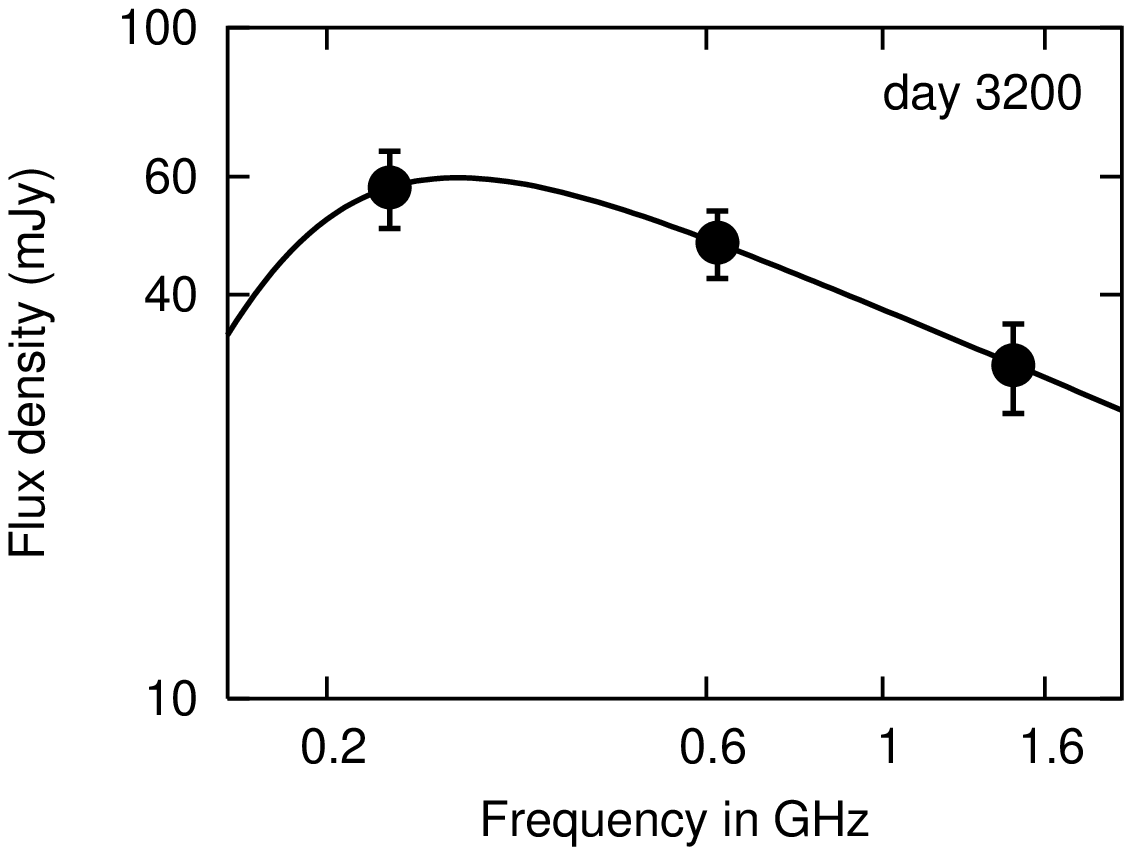}\\
  \plotone{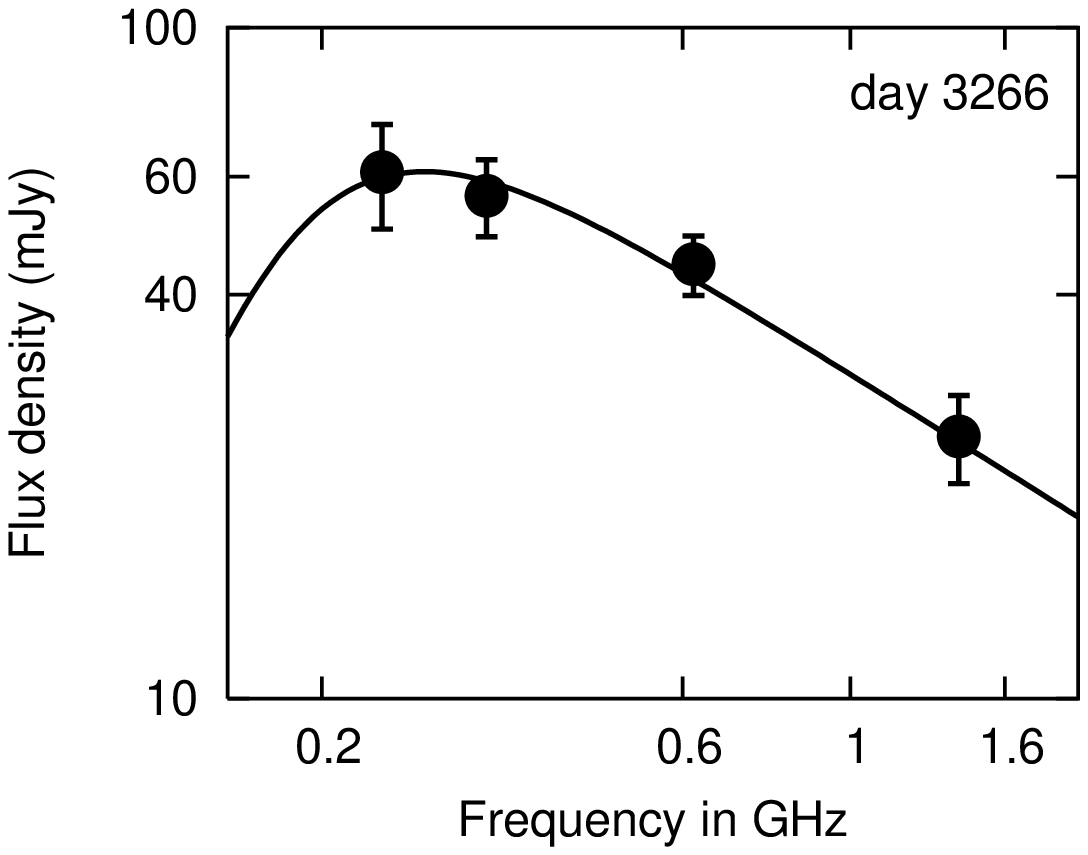}\hfil\plotone{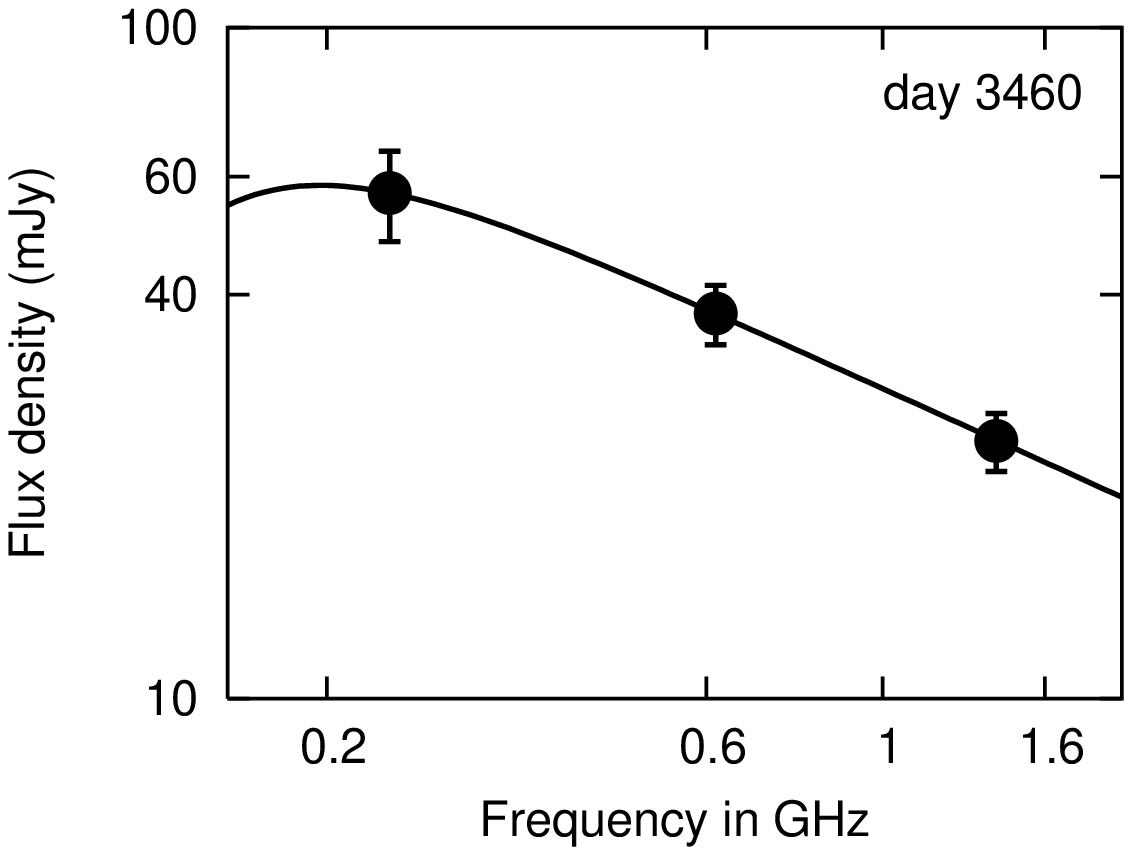}\\
  \plotone{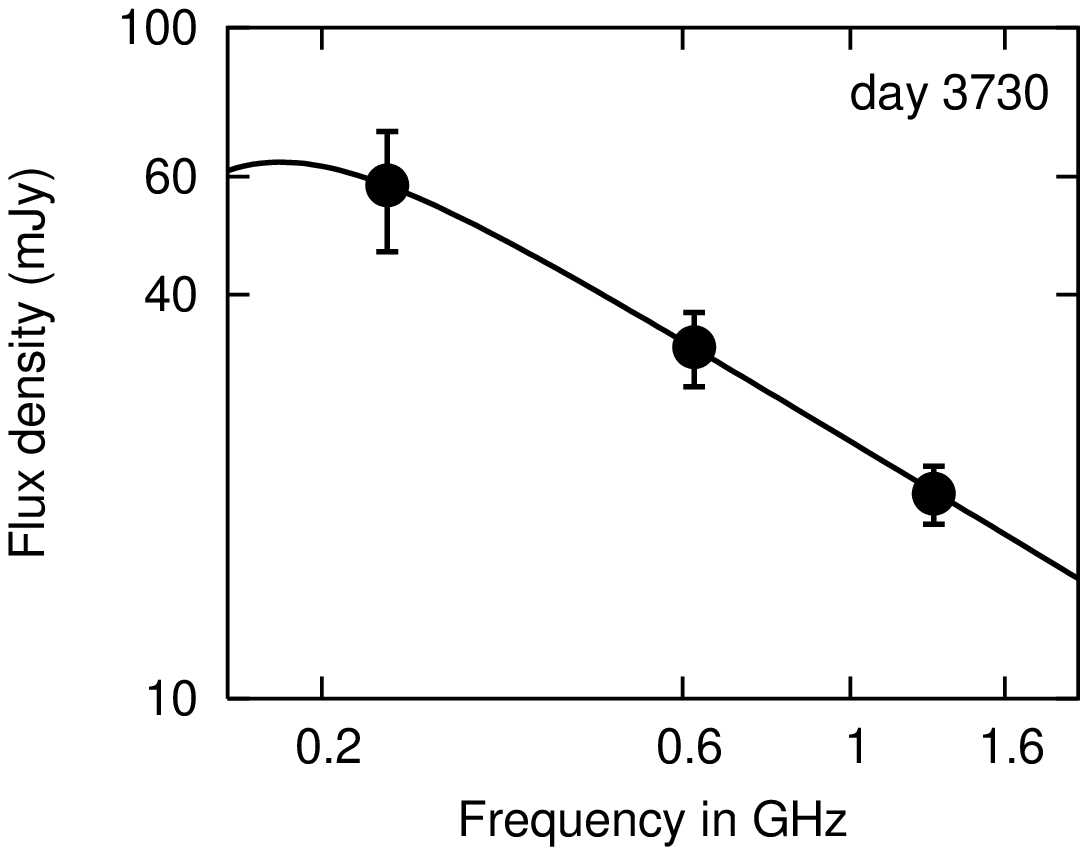}
\caption{Best fit GMRT spectra of SN 1993J on 
day $\sim 3000$, 3200, 3266, 3460 and 3730 since explosion from top left
in clockwise order. Solid lines are the 
SSA best fit models (See Table \ref{bestfit} for details).
}
\label{spectra1}
\end{figure}

\clearpage

\begin{figure}
\epsscale{1}
\plotone{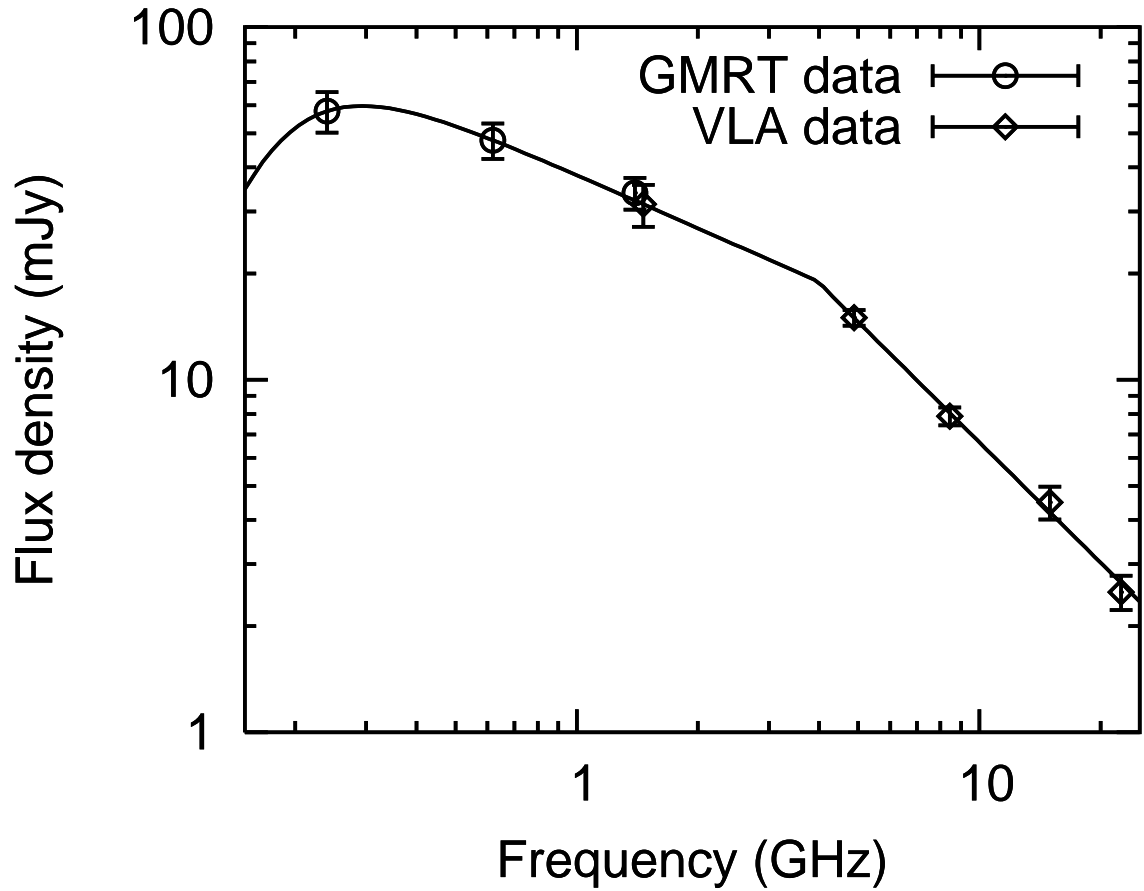}
\caption{GMRT plus VLA combined spectrum on day 3200. Solid line
is the best fit SSA model. The spectrum shows a 
break at 4 GHz with steepening in the spectral index by 0.62 \citep{cha04}.}
\label{comb}
\end{figure}

\clearpage

\begin{figure}
\epsscale{1}
\plotone{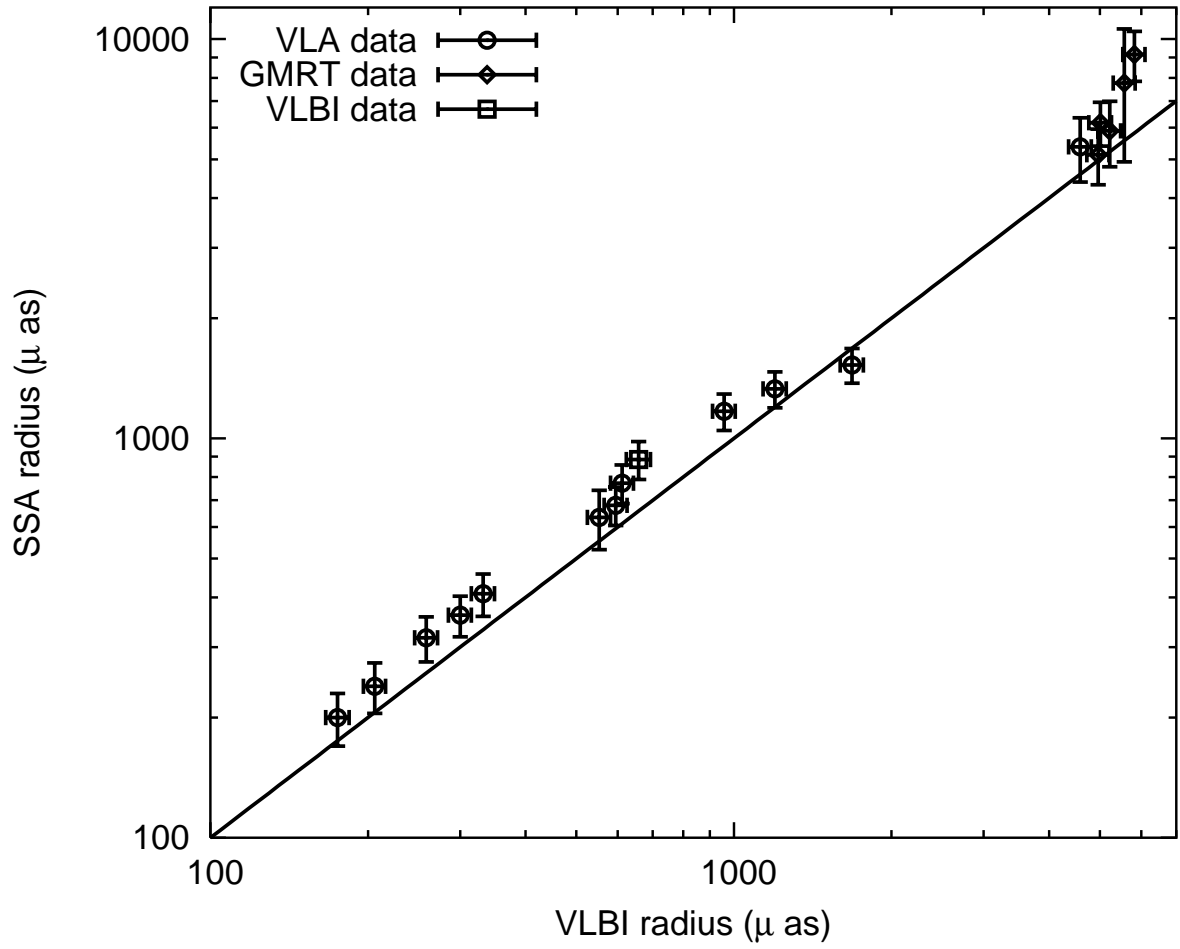}
\caption{Plot of the size of SN 1993J from SSA fit against
the VLBI size. The straight line indicates the region where the 
SSA size equals the VLBI size of the supernova.}
\label{size}
\end{figure}

\clearpage

\begin{figure}
\epsscale{1}
\plotone{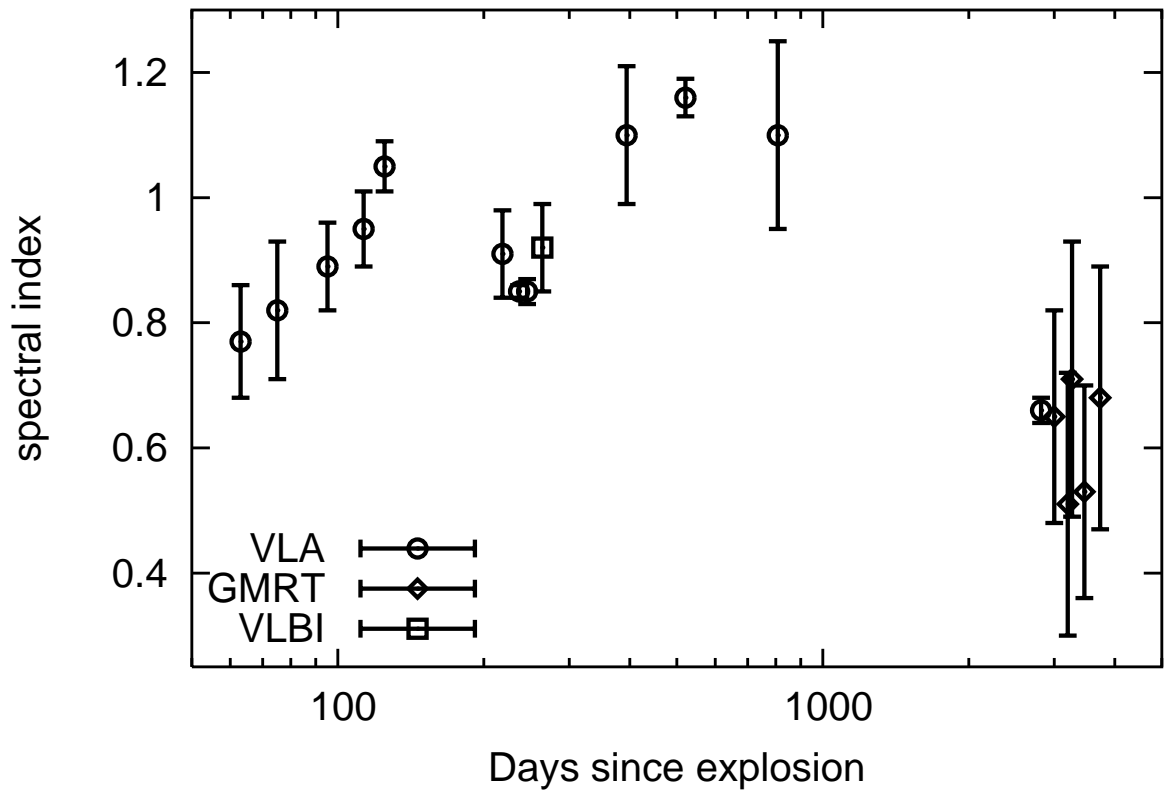}
\caption{Evolution of the best
fit spectral index $\alpha$ with time. Spectral index changes from $\sim 1$ 
to $\sim 0.6$.}
\label{alpha.mag}
\end{figure}
\clearpage

\begin{figure}
\begin{center}
\includegraphics[scale=.50,angle=270]{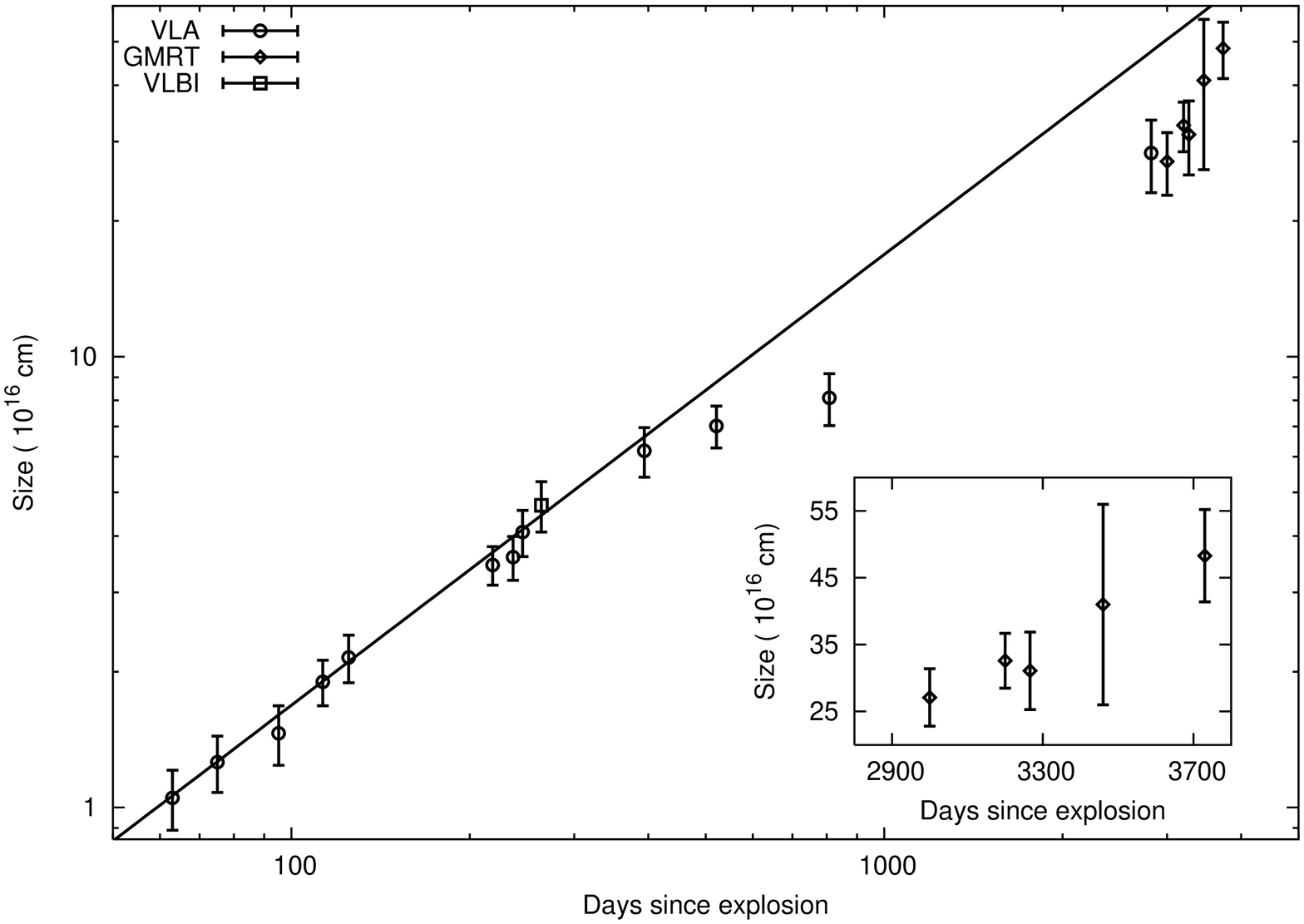}
\includegraphics[scale=.50,angle=270]{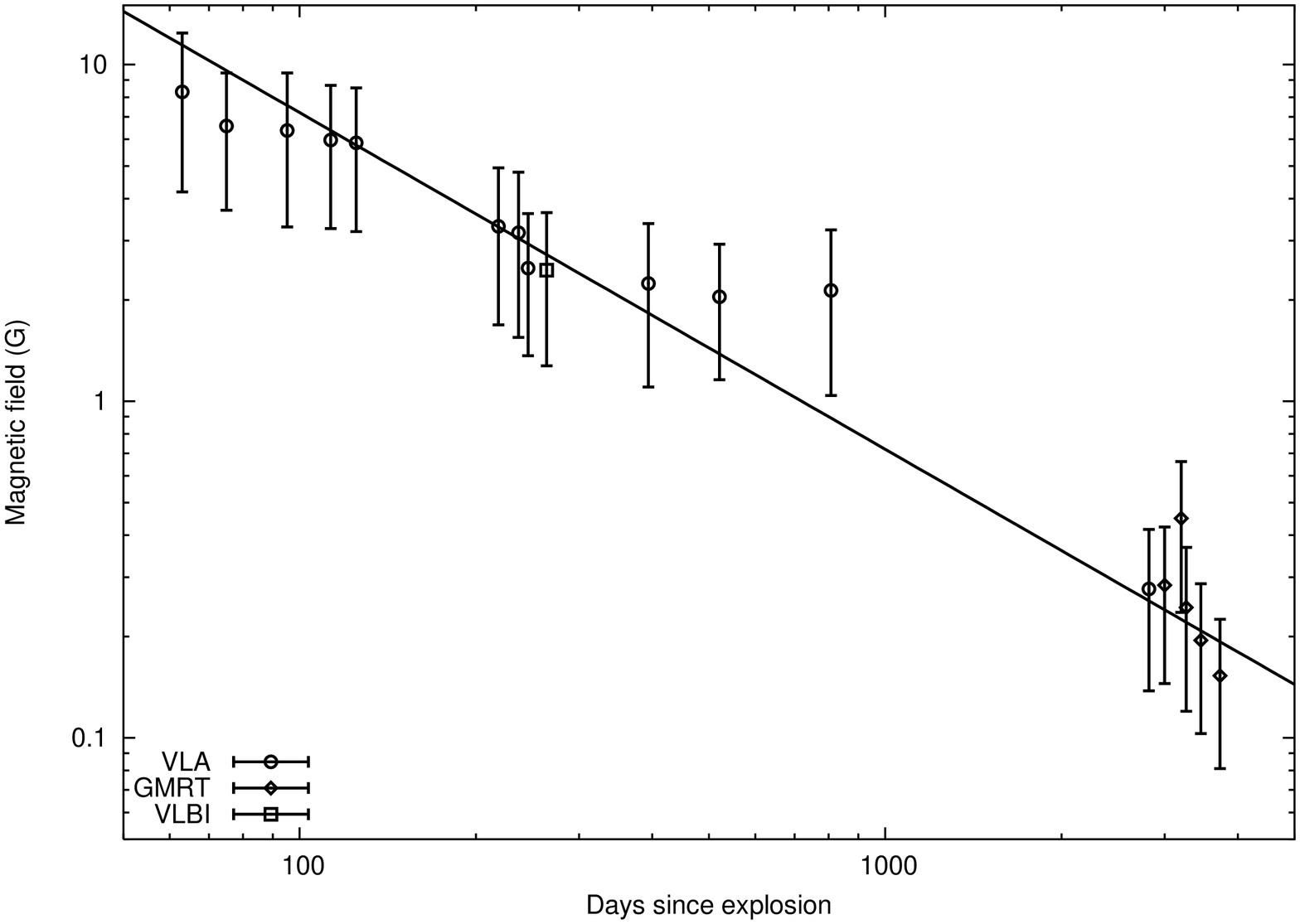}
\end{center}
\caption{Evolution of the size of the supernova with time (upper
  panel). The straight line corresponds to $R \propto t $ i.e. free
  expansion. Initially the expansion is free expansion but later it
  shows a deceleration. The inset shows the evolution of size
  determined only from GMRT data (note the linear scales in time and
  radius here). Evolution of the magnetic field in synchrotron
  emitting plasma with time (lower panel).  Plot also shows the line
  corresponding to $B \propto t^{-1}$.  The magnetic field evolution
  is roughly consistent with $B \propto t^{-1}$ though it shows some
  possible flattening around day 300-900.} 
\label{size.mag}
\end{figure}

\clearpage

\begin{figure}
\plotone{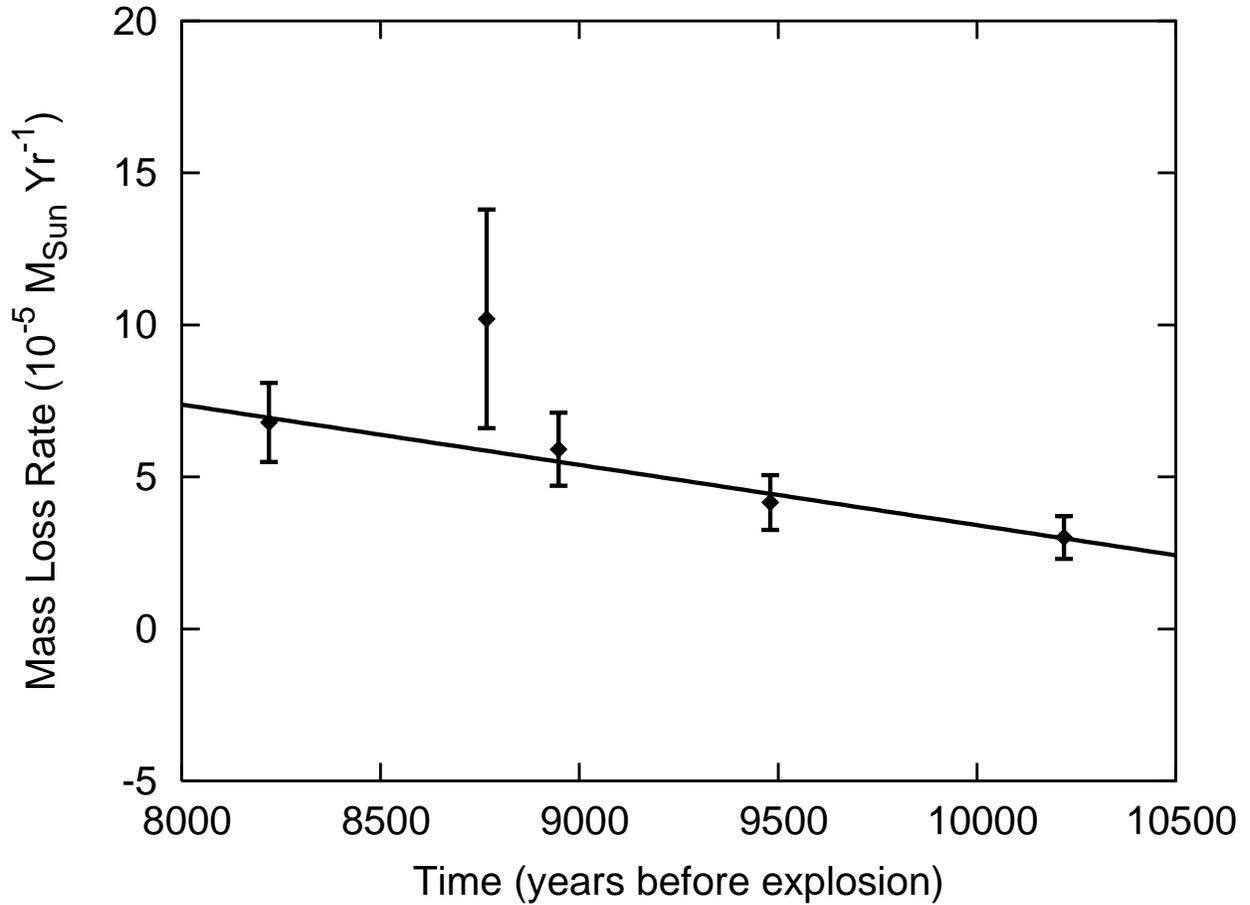}
\caption{Evolution of the mass loss rate of the presumed red supergiant
progenitor of SN 1993J with time before explosion of the star.
The solid line represents the best fit with a slope 
$1.8 \times 10^{-8} {\rm {M}_{\odot}\, {yr}^{-1}/yr}$. 
Here the ratio of the ejecta velocity and wind velocity is taken to be
$\sim 1000$ to compute the corresponding time before explosion.}
\label{mass}
\end{figure}
\clearpage

\begin{deluxetable}{llrrccccc}
\tablewidth{480pt}
\tablecaption{Details of the flux calibrators used
\label{fluxcal}}
\tablehead{\colhead{} & \colhead{} & \multicolumn{2}{c}{Co-ordinates(J2000)} &
\colhead{} & \multicolumn{4}{c}{Flux Density (Jy)\tablenotemark{a}}\\
\cline{3-4} \cline{6-9} 
\colhead{Name} & \colhead{Redshift} & \colhead{RA} & \colhead{DEC}& 
\colhead{}
& \colhead{1420MHz} & \colhead{610MHz} & \colhead{325MHz}& \colhead{235MHz}}
\startdata
3C48 & 0.367 & $01^{\mathrm{h}}37^{\mathrm{m}}41\fs3$ &
$33\arcdeg09\arcmin35\arcsec$ & &  16.2 &  29.4 & 43.4 & 50.8\\
3C147 & 0.545 & $05^{\mathrm{h}}42^{\mathrm{m}}36\fs1$ &
$49\arcdeg51\arcmin07\arcsec$ & &  22.3 & 38.3& 52.7 & 59.2\\
3C286 & 0.849 & $13^{\mathrm{h}}31^{\mathrm{m}}08\fs3$ &
$30\arcdeg30\arcmin33\arcsec$ & &  $-$ &  21.1 & 25.9 & 28.1\\
\enddata
\tablenotetext{a}{
The flux density of the flux calibrators were calculated using Baars 
formulation \citep{bar77}. Here it was assumed that these sources
evolve as single power law with respect to 
frequencies, taken from VLA calibrator
catalogue.}
\end{deluxetable}

\clearpage 

\begin{deluxetable}{llccc}
\tablewidth{335pt}
\tablecaption{GMRT observations of the phase calibrator 1035+564 at 1390 MHz band
\label{1035cal}}
\tablehead{\colhead{Date of} & \colhead{Flux Density} & 
\multicolumn{2}{c}{Observed position J2000} & \colhead{Offset\tablenotemark{c}}\\
\cline{3-4} 
\colhead{Observn.} & \colhead{Jy$\pm$ mJy\tablenotemark{a}}&
\colhead{RA} & \colhead{DEC} & \colhead{arcsec} }
\startdata 
2001 Oct 15 & $2.46\pm3.9$ & $10^{\mathrm{h}}35^{\mathrm{m}}07\fs1$ & 
$56\arcdeg28\arcmin47\arcsec$ & 1.5\arcsec\\
2002 Apr 07 & $2.39\pm5.6$ & $10^{\mathrm{h}}35^{\mathrm{m}}07\fs0$ &
$56\arcdeg28\arcmin48\arcsec$ & 1.0\arcsec \\
2002 Jun 24& $2.08\pm2.9$ & $10^{\mathrm{h}}35^{\mathrm{m}}06\fs9$ &
$56\arcdeg28\arcmin47\arcsec$ & 1.5\arcsec \\
2002 Sep 21 & $2.33\pm3.3$ & $10^{\mathrm{h}}35^{\mathrm{m}}07\fs1$ &
$56\arcdeg28\arcmin46\arcsec$ & 1.8\arcsec \\
2003 Jun 13\tablenotemark{b} & $1.89\pm1.6$ &
$10^{\mathrm{h}}35^{\mathrm{m}}07\fs0$ & $56\arcdeg28\arcmin47\arcsec$ &
0.0\arcsec \\
\enddata
\tablenotetext{a}{Note that the flux density is in Jy and error in mJy. Errors are
best fit errors obtained from AIPS task JMFIT.}
\tablenotetext{b}{Note that this observation is at frequency 1280 MHz whereas
other observations are at 1390 MHz.} 
\tablenotetext{c}{ Offsets are from the optical position of the sources 
given in NED.}
\end{deluxetable}

\clearpage

\begin{deluxetable}{llrccc}
\tablewidth{405pt}
\tablecaption{Details of the GMRT observations of phase calibrator 0834+555
\label{phasecal}}
\tablehead{\colhead{Date of} & \colhead{Frequency} & 
\colhead{Flux Density} & 
\multicolumn{2}{c}{Observed position J2000} & \colhead{Offset\tablenotemark{b}}\\
\cline{4-5} 
\colhead{Observn.} & \colhead{MHz} & \colhead{Jy$\pm$ mJy\tablenotemark{a}} &
\colhead{RA} & \colhead{DEC} & \colhead{arcsec} }
\startdata 
2000 Nov 08 & 1420 & $8.80\pm8.5$ & $08^{\mathrm{h}}34^{\mathrm{m}}54\fs9$ & 
$55\arcdeg34\arcmin24\arcsec$ & 3.0\arcsec\\
2000 Dec 16& 1420 & $8.75\pm6.6$ & $08^{\mathrm{h}}34^{\mathrm{m}}55\fs0$ & 
$55\arcdeg34\arcmin14\arcsec$ & 7.1\arcsec\\
2001 Jun 02& 1390 & $8.71\pm4.2$ & $08^{\mathrm{h}}34^{\mathrm{m}}52\fs2$ & 
$55\arcdeg34\arcmin38\arcsec$ & 43.9\arcsec\\
\hline
2001 Mar 24& 610 & $8.47\pm17.8$ & $08^{\mathrm{h}}34^{\mathrm{m}}54\fs9$ & 
$55\arcdeg34\arcmin34\arcsec$ & 13.0\arcsec\\
2001 Aug 24& 610 & $8.18\pm12.6$ & $08^{\mathrm{h}}34^{\mathrm{m}}55\fs1$ & 
$55\arcdeg34\arcmin20\arcsec$ & 1.8\arcsec\\
2001 Dec 30& 610 & $8.74\pm13.8$ & $08^{\mathrm{h}}34^{\mathrm{m}}54\fs9$ & 
$55\arcdeg34\arcmin22\arcsec$ & 1.0\arcsec\\
2002 Mar 08& 610 & $8.23\pm12.5$ & $08^{\mathrm{h}}34^{\mathrm{m}}54\fs9$ & 
$55\arcdeg34\arcmin21\arcsec$ & 0.0\arcsec\\
2002 May 19& 610 & $8.44\pm20.0$ & $08^{\mathrm{h}}34^{\mathrm{m}}54\fs9$ & 
$55\arcdeg34\arcmin21\arcsec$ & 0.0\arcsec\\
2002 Sep 16& 610 & $8.50\pm11.3$ & $08^{\mathrm{h}}34^{\mathrm{m}}54\fs9$ & 
$55\arcdeg34\arcmin21\arcsec$ & 0.0\arcsec\\
2003 Jun 17& 610 & $8.14\pm15.7$ & $08^{\mathrm{h}}34^{\mathrm{m}}54\fs9$ & 
$55\arcdeg34\arcmin21\arcsec$ & 0.0\arcsec\\
\hline
2001 Jul 05& 325 & $8.96\pm26.6$ & $08^{\mathrm{h}}34^{\mathrm{m}}55\fs0$ & 
$55\arcdeg34\arcmin24\arcsec$ & 3.4\arcsec\\
2002 Mar 07& 325 & $8.76\pm6.8$ & $08^{\mathrm{h}}34^{\mathrm{m}}55\fs0$ & 
$55\arcdeg34\arcmin22\arcsec$ & 1.8\arcsec\\
2002 Aug 16& 325 & $9.06\pm31.9$ & $08^{\mathrm{h}}34^{\mathrm{m}}54\fs9$ & 
$55\arcdeg34\arcmin21\arcsec$ & 0.0\arcsec\\
\hline
2001 Dec 31& 243 & $8.77\pm46.8$ & $08^{\mathrm{h}}34^{\mathrm{m}}54\fs9$ & 
$55\arcdeg34\arcmin21\arcsec$ & 0.0\arcsec\\
2002 Mar 08& 243 & $8.20\pm36.6$ & $08^{\mathrm{h}}34^{\mathrm{m}}55\fs0$ & 
$55\arcdeg34\arcmin24\arcsec$ & 3.4\arcsec\\
2002 Sep 16& 243 & $9.02\pm43.8$ & $08^{\mathrm{h}}34^{\mathrm{m}}55\fs0$ & 
$55\arcdeg34\arcmin20\arcsec$ & 1.8\arcsec\\
2003 Jun 17& 243 & $8.25\pm44.7$ & $08^{\mathrm{h}}34^{\mathrm{m}}54\fs9$ & 
$55\arcdeg34\arcmin21\arcsec$ & 0.0\arcsec\\

\enddata
\tablenotetext{a}{Note that the flux is in Jy and error in mJy. Errors are
best fit errors obtained from AIPS task JMFIT}
\tablenotetext{b}{ Offsets are from the optical position of the sources
given in NED.}

\end{deluxetable}

\clearpage

\begin{deluxetable}{lcccccc}
\tabletypesize{\footnotesize}
\tablewidth{420pt}
\tablecaption{Details of the GMRT observations of SN 1993J
\label{tab:1}}
\tablehead{\colhead{Date of}   & \colhead{Days since} & \colhead{Freq. Band}
& \colhead{No. of good} &\colhead{ Resolution} & \colhead{Flux density}
 & \colhead{rms} \\
\colhead{Observn} & \colhead{explosion}  & \colhead{MHz}    &  
\colhead{Antennas} &\colhead{arcsec} & \colhead{mJy}     & \colhead{mJy} } 
\startdata
2000 Nov 08& 2779 & 1420 &19  & 25x18 & $35.1\pm3.5$   & 0.5\\
2000 Dec 16& 2818 & 1420 &23  & 8x5   & $36.1\pm3.6$   & 0.3\\
2001 Jun 02& 2988 & 1390 &28  & 4x3   & $32.7\pm3.3$   & 0.2\\
2001 Oct 15& 3123 & 1390 &24  & 10x6  & $33.9\pm3.3$   & 0.3\\
2002 Apr 07& 3296 & 1390 &25  & 11x7  & $24.6\pm3.7$   & 1.0\\
2002 Jun 24& 3374 & 1390 &19  & 6x3   & $23.4\pm2.5$   & 0.4\\
2002 Sep 21& 3463 & 1390 &25  & 5x3   & $24.2\pm2.4$   & 0.2\\
2003 Jun 13& 3728 & 1280 &24  & 5x2   & $20.2\pm2.1$   & 0.2\\
\hline
2001 Mar 24& 2917 & 610 &20  & 11x7  & $56.1\pm5.5$   & 0.5\\
2001 Aug 24& 3072 & 610 &24  & 13x7  & $55.8\pm5.7$   & 0.4\\
2001 Dec 30& 3198 & 610 &20  & 11x8  & $47.8\pm5.5$   & 1.9\\
2002 Mar 08& 3266 & 610 &25  & 14x6  & $44.4\pm4.5$   & 0.3\\
2002 May 19& 3338 & 610 &24  & 18x8  & $44.6\pm4.5$   & 0.6\\
2002 Sep 16& 3458 & 610 & 26  & 10x6  & $37.5\pm3.8$  & 0.4\\
2003 Jun 17& 3732 & 610 &23  & 14x5  & $33.4\pm4.3$   & 0.8\\
\hline
2001 Jul 05& 3022 & 325 &18  & 19x10 & $69.2\pm15.8$ & 2.5\\
2002 Mar 07& 3265 & 325  &24 & 19x13 & $55.8\pm7.4$   & 1.9\\
2002 Aug 16& 3427 & 325  &16 & 18x9  & $61.8\pm8.8$   & 2.7\\
\hline
2001 Dec 31& 3199 & 243 &20  & 19x14 & $57.8\pm7.6$   & 2.5\\
2002 Mar 08& 3266 & 243 &17  & 26x16 & $60.9\pm10.8$  & 4.1\\
2002 Sep 16& 3458 & 243 &17  & 26x16 & $56.7\pm8.7$   & 4.0\\
2003 Jun 17& 3732 & 243 &22  & 23x11 & $58.2\pm11.8$  & 5.4\\
\enddata
\end{deluxetable}

\clearpage

\begin{deluxetable}{llccc}
\tabletypesize{\footnotesize}
\tablewidth{290pt}
\tablecaption{Details of the near simultaneous spectra of SN 1993J
\label{tab:2}}
\tablehead{\colhead{Date of} & \colhead{Days since} & \colhead{Freq. Band}
& \colhead{Flux density} 
& \colhead{rms}\\
\colhead{Obs.}   & \colhead{explosion}  & \colhead{ MHz}    &  
 \colhead{mJy}  & \colhead{mJy}} 
\startdata
2001 Jul 05  & 3022 & 325   & $69.2\pm15.8$  & 2.5\\
2001 Aug 24 & 3072 & 610    & $55.8\pm5.7$   & 0.4\\
2001 Jun 02  & 2988 & 1390  & $32.7\pm3.3$  & 0.2\\
\hline
2001 Dec 31 & 3199 & 243    & $57.8\pm7.6$   & 2.5\\
2001 Dec 30 & 3198 & 610    & $47.8\pm5.5$   & 1.9\\
2001 Oct 15                  & 3123 & 1390 & $33.9 \pm3.5$   & 0.3\\
2002 Jan 13\tablenotemark{a} & 3212 & 1460  & $31.44\pm4.28$  & 2.9\\
2002 Jan 13\tablenotemark{a} & 3212 & 4885  & $15   \pm0.77$  & 0.19\\
2002 Jan 13\tablenotemark{a} & 3212 & 8440   & $7.88 \pm0.46$  & 0.24\\
2002 Jan 13\tablenotemark{a} & 3212 & 14965 & $4.49 \pm0.48$  & 0.34\\
2002 Jan 13\tablenotemark{a} & 3212 & 22485 & $2.50 \pm0.28$  & 0.13\\

\hline
2002 Mar 08 & 3266 & 243 & $60.9\pm10.8$  & 4.1\\
2002 Mar 07 & 3265 & 325 & $56.2\pm7.4$   & 1.9\\
2002 Mar 08 & 3266 & 610 & $44.4\pm4.5$   & 0.3\\
2002 Apr 07 & 3296 & 1390 & $24.6\pm3.7$   & 1.0\\
\hline
2002 Sep 16& 3458 & 243  & $56.7\pm8.7$   & 4.0\\
2002 Sep 16 & 3458 & 610  & $37.5\pm3.8$   & 0.4\\
2002 Sep 21 & 3463 & 1390 & $24.2\pm2.4$   & 0.2\\
\hline
2003 Jun 17 & 3732 & 243  & $58.2\pm11.8$ & 5.4\\
2003 Jun 17 & 3732 & 610  & $33.4\pm4.3$   & 0.8\\
2003 Jun 13 & 3728 & 1280 & $20.2\pm2.1$   & 0.2\\
\enddata
\tablenotetext{a}{VLA data, courtesy K. Weiler and collaboration }
\end{deluxetable}

\clearpage

\begin{deluxetable}{llccc}
\tablewidth{310pt}
\tablecaption{Best fit parameters of SSA fits to the GMRT 
spectra 
\label{bestfit}}
\tablehead{\colhead{Days since} & \colhead{ Spectral index } &
\multicolumn{2}{c}{SSA best fit}\\
\cline{3-4} 
\colhead{explosion} & \colhead{$\alpha$\tablenotemark{a} } & \colhead{$B$ Gauss} &
\colhead{$R \times 10^{17}$ cm} & \colhead{}}
\startdata
3000 & $0.65\pm0.17$ & $0.28\pm0.12$& $2.71\pm0.43$\\ 
3200 & $0.51\pm0.21$\tablenotemark{b} & $0.43\pm0.19$\tablenotemark{c}
& $3.26\pm0.41$\\
3266 & $ 0.71\pm0.22$ & $0.24\pm0.11$& $3.11\pm0.58$\\
3460 & $0.53\pm0.17$ & $0.19\pm0.09$& $4.10\pm0.50$\\
3730 & $0.68\pm0.21$ & $0.15\pm0.07$& $4.83\pm0.69$\\

\enddata
\tablenotetext{a}{$\alpha$ is calculated manually from the
assumed optically thin part of the spectrum between 610 and 1420 MHz.}
\tablenotetext{b}{Spectral index $\alpha$ {\it before} the
 break in the spectrum.}
\tablenotetext{c}{This determination of $B$ is from the best fit SSA
model. The  magnetic field using synchrotron cooling break is 0.33 G. Both
values match within error bars.}
\end{deluxetable}

%
\clearpage
\begin{deluxetable}{llc}
\tablewidth{240pt}
\tablecaption{Lower limits to the mass loss rates of the
progenitor of SN 1993J from Eq. \ref{massloss} with m=0.781
\label{mass_loss}}
\tablehead{\colhead{Days since} & \colhead{ Years before } &
 \colhead{ Mass Loss Rate}\\
\colhead{explosion} & \colhead{explosion\tablenotemark{a} }
& \colhead{$10^{-5}\, {\rm M_{\odot} Yr^{-1}}$ } }
\startdata
3000 & 8219 & $6.8\pm1.3$\\ 
3200 & 8767 & $10.2\pm3.6$\\ 
3266 & 8948 & $5.9\pm1.2$\\ 
3460 & 9480 & $4.2\pm0.9$\\
3730 & 10219 & $3.01\pm0.7$\\
\enddata
\tablenotetext{a}{Years before explosion is calculated by taking
the ratio of the ejecta velocity to the wind velocity to be
$\sim 1000$.}
\end{deluxetable}

\end{document}